\documentclass[a4paper,11pt]{article}
\pdfoutput=1 

\usepackage{jheppub} 

\usepackage[T1]{fontenc} 
\usepackage[numbers]{natbib}
\usepackage{filecontents}
\usepackage{dynkin-diagrams}
\usepackage{tikz}
\usepackage{float}
\usepackage{placeins}
\usepackage{stackengine}
\tikzset{Rightarrow/.style={double equal sign distance,>={Implies},->},
triple/.style={-,preaction={draw,Rightarrow}},
quadruple/.style={preaction={draw,shorten >=0pt},shorten >=1pt,-,double,double
distance=0.2pt}}
\usetikzlibrary{positioning}
\usepackage{graphicx}
\usepackage{pdflscape}
\usepackage{caption}
\usepackage{subcaption}
\usepackage{comment}
\usepackage{subfiles}
\usepackage{todonotes}
\usepackage{booktabs}

\usetikzlibrary{hobby,intersections} 
\usetikzlibrary{shapes.geometric}
\usetikzlibrary{shapes.misc}

\tikzset{flavour/.style={draw=none,minimum size=0.3mm,fill=white, regular polygon,regular polygon sides=4,draw}}
\tikzset{flavourr/.style={draw=none,minimum size=0.3mm,fill=red, regular polygon,regular polygon sides=4,draw}}
\tikzset{gaugeBig/.style={inner sep=1mm,draw=none,fill=white,minimum size=2mm,circle, draw}}
\tikzset{bd/.style={circle, draw=black, inner sep=0pt, fill=black, minimum size=2mm}}
\tikzset{wd/.style={circle, draw=black, inner sep=0pt, fill=white, minimum size=2mm}}
\tikzset{Dynkin/.style={circle, draw=black, inner sep=0pt, fill=white, minimum size=2mm}}
\tikzstyle{ligne}=[draw, very thick] 
\tikzstyle{gridline}=[draw, gray] 
\tikzset{gauge/.style={circle, draw,inner sep=2.5pt}}
\tikzset{gaugeo/.style={circle, draw,inner sep=2.5pt,fill=orange}}
\tikzset{gauger/.style={circle, draw,inner sep=2.5pt,fill=red}}
\tikzset{gaugeb/.style={circle, draw,inner sep=2.5pt,fill=blue}}
\tikzset{gaugeg/.style={circle, draw,inner sep=2.5pt,fill=green}}
\tikzset{gaugegoodgreen/.style={circle, draw,inner sep=2.5pt,fill=goodgreen}}
\tikzset{gaugem/.style={circle, draw,inner sep=2.5pt,fill=magenta}}
\tikzset{hasse/.style={circle, fill,inner sep=2pt}}
\tikzset{d2/.style={circle, fill,inner sep=1.3pt}}
\tikzset{shrinky/.style={circle, fill,inner sep=1pt}}
\tikzset{sized/.style={circle, draw, inner sep=1.5pt}}
\tikzset{seven/.style={circle, draw,inner sep=3pt}}

\makeatletter
\DeclareRobustCommand{\rvdots}{%
  \vbox{
    \baselineskip4\p@\lineskiplimit\z@
    \kern-\p@
    \hbox{.}\hbox{.}\hbox{.}
  }}
\makeatother

\newcommand{\Figref}[1]{Figure~\ref{#1}}
\newcommand{\Quiver}[1]{$\mathcal Q_{\ref{#1}}$}
\newcommand{\surm}{\mathrm{SU}}

\newcommand{\urm}{\mathrm{U}}
\newcommand{\sorm}{\mathrm{SO}}
\newcommand{\orm}{\mathrm{O}}
\newcommand{\sprm}{\mathrm{Sp}}
\newcommand{\hs}{\mathrm{HS}}

\newcommand{\hwg}{\mathrm{HWG}}
\newcommand{\pe}{\mathrm{PE}}
\newcommand{\pl}{\mathrm{PL}}

\title{\boldmath New symplectic singularities from $\surm(2)$ gauge theories}

\author[a]{Amihay Hanany, Guhesh Kumaran, Deshuo Liu,}
\author[b]{and Travis Schedler}

\affiliation[a]{Abdus Salam Centre for Theoretical Physics, Imperial College London,\\ Prince Consort Road
London, SW7 2AZ, UK}
\affiliation[b]{Department of Mathematics, Imperial College London,\\ Prince Consort Road
London, SW7 2AZ, UK}

\emailAdd{a.hanany@ic.ac.uk}
\emailAdd{guhesh.kumaran18@imperial.ac.uk}
\emailAdd{deshuo.liu21@imperial.ac.uk}
\emailAdd{t.schedler@imperial.ac.uk}

\preprint{Imperial/TP/26/AH/06}
\abstract{Families of $\sprm(1)\simeq\surm(2)$ gauge theories with eight supercharges are found to have a Higgs branch which is an isolated symplectic singularity. These are, in some sense, the most ``minimal'' gauge theories as they only Higgs to a trivial theory. The matter content is $N$ fundamental half-hypermultiplets and one half-hypermultiplet in the $\mathrm{Sym}^k$ representation where $k=1,3,5,7$. The cases of $k=1,3$ reproduce the known Kraft--Procesi construction for minimal nilpotent orbit closures of $\sorm(N+1)$ and the $g\sorm(N)$ singularities of \cite{Bourget:2025wsp}, respectively. The cases $k=5,7$ are new isolated symplectic singularities which are termed $h\sorm(N)$ and $i\sorm(N)$, respectively. The classification of these isolated symplectic singularities is argued for through the Higgs mechanism, with Hilbert series and highest weight generating (HWG) functions computed for some cases. Each of the $g\sorm(N)$, $h\sorm(N)$, and $i\sorm(N)$ families has 
a (quaternionic) one-dimensional member; these
are the Klein $A_3$, $E_6$, and $E_8$ singularities, respectively. Our construction hence provides realisations of these Klein singularities  
as Higgs branches (hyper-Kähler quotients) of $\sprm(1)$ gauge theories, 
complementary to Kronheimer's construction 
\cite{Kronheimer:1989zs} using the $\widehat A_3$, $\widehat E_6$, and $\widehat E_8$ affine quivers. The Klein $E_7$ singularity is also realised as a Higgs branch (hyper-Kähler quotient) of an $\sprm(1)\times\orm(1)$ gauge theory.}
\begin{document}
\maketitle
\flushbottom
\section{Introduction}
The moduli spaces of vacua of $3d\;\mathcal N=4$ theories have two distinct branches called the Higgs branch and the Coulomb branch. 
To physicists, these branches are hyper-Kähler cones, of which many examples are symplectic singularities \cite{2000InMat.139..541B}.
To mathematicians, these branches are Poisson varieties 
which enjoy a finite stratification by symplectic leaves, partially ordered by inclusion of their closures.
This partial order is captured in a Hasse diagram which encodes important information about the gauge theory and the structure of the variety. The symplectic leaves of the Higgs and Coulomb branches have different physical interpretations. On the Higgs branch, the different symplectic leaves correspond to the different effective field theories that arise from Higgsing the theory through the Brout--Englert--Higgs--Guralnik--Hagen--Kibble mechanism (or ``Higgs mechanism'' for short) \cite{Englert:1964,Higgs:1964,Guralnik:1964,Kibble:1967}. These effective field theories appear in a ``bottom-up'' way in the Hasse diagram through successive 
Higgsings. Similarly, there are associated gauge theories for the symplectic leaves on the Coulomb branch. The interpretation of the gauge theory for symplectic leaves on the Coulomb branch is as a description of the massless states that are preserved at different points on the Coulomb branch. The theories associated to symplectic leaves on the Coulomb branch appear in a ``top-down'' way in the Hasse diagram.

The simplest possible stratification occurs in an isolated symplectic singularity, which has only two symplectic leaves: one consists of the singular point and the other of its smooth complement. These are, in a sense, building blocks of general symplectic singularities and often serve as inputs into the algorithms mentioned above. These isolated symplectic singularities are also interesting to physicists outside of the context of Higgsing gauge theories. For example, the $ADE$ Klein singularities are some of the most common string backgrounds and minimal nilpotent orbit closures are more widely known to physicists as centred one-instanton
moduli spaces 
on $\mathbb C^2$.

A list of isolated symplectic singularities which appear in the Hasse diagram of the magnetic spectrum of $3d\;\mathcal N=4$ quiver gauge theories with unitary gauge groups was proposed in \cite{Bourget:2025wsp} using the ``decay and fission'' algorithm. In this list, a new one-parameter family of isolated singularities
was discovered via 
magnetic quivers, that is, Coulomb branch constructions,  
and named
$gb_n$ and $gd_n$; these shall be referred to collectively as $g\sorm(N)$ here for odd and even $N$ respectively. The term ``magnetic'' refers to the construction of the Coulomb branch as a moduli space of dressed monopole operators which are magnetic objects \cite{Cremonesi:2013lqa}. 

In this paper, a hyper-Kähler quotient construction of the $g\sorm(N)$ singularity is presented using the technique of chain polymerisation \cite{Hanany:2024fqf} on magnetic quivers for moduli spaces of free fields. This hyper-Kähler quotient construction yields
an electric quiver, i.e. a Higgs branch construction, for the $g\sorm(N)$ singularity involving an $\sprm(1)\simeq\surm(2)$ gauge theory with $\mathrm{Sym}^3$ matter and $N$ half-hypermultiplets. The term ``electric'' refers to the Higgs branch being a moduli space of gauge invariants constructed from electrically charged hypermultiplets. The electric quiver is then generalised and two more families of isolated singularities named $h\sorm(N)$ and $i\sorm(N)$ are found which replace the $\mathrm{Sym}^3$ matter with $\mathrm{Sym}^5$ and $\mathrm{Sym}^7$, respectively. A proof that these are the only other isolated singularities from this construction is also given by analysing the Higgsing of these theories.

The electric quivers permit computation of a Higgs branch Hilbert series, corresponding to the isolated symplectic singularities, and also the Coulomb branch Hilbert series, thereby proposing candidate symplectic duals to these singularities. This is in contrast with the magnetic quivers for the $g\sorm(N)$ singularities which preclude computation of Higgs branch Hilbert series 
owing to non-simply laced edges. The electric quivers also elucidate gauge anomalies for anomalous choices
of $k$ and $N$
that
are not apparent in the magnetic quiver. Nevertheless, the construction of symplectic singularities is agnostic to any anomalies.

With the electric quivers it is also possible to compute the (quaternionic) dimension one singularities which are the lowest dimension members of the $g\sorm(N),\;h\sorm(N),$ and $i\sorm(N)$ families. These are the Klein $A_3,\;E_6,$ and $E_8$ singularities, respectively, which gives a novel hyper-Kähler quotient construction of these singularities as the Higgs branch of an $\sprm(1)$ gauge theory and complements the construction of Kronheimer \cite{Kronheimer:1989zs} as the Higgs branch of the respective affine quivers.

\paragraph{Note on AI usage} Claude Opus 4.8 was used to generate a Python script which was then used by a human to compute unrefined Hilbert series only for the cases with $\mathrm{Sym}^7$. The computations were independently checked by a human to some orders. 
ChatGPT 5.5 and 5.6 were used to generate parts of Appendix \ref{app_branching} and to make some final proofreading suggestions, which were reviewed by a human.

\paragraph{Organisation of the paper} 
Section \ref{sec:chain} applies the chain polymerisation method to construct the $g\sorm(N)$ quivers of \cite{Bourget:2025wsp} and identifies
the resulting hyper-Kähler quotient construction.

In Section \ref{sec:elecquiver}, the two-parameter family of electric quivers is introduced. An argument is made through the Higgs mechanism that,
beyond the familiar $k=1$ case, only $\mathrm{Sym}^k$ matter with $k=3,5,7$ produces further isolated symplectic singularities. These are called $g\sorm(N),\;h\sorm(N),$ and $i\sorm(N)$, respectively.

In Section \ref{sec:HiggsBranchHS}, 
the Higgs branch Hilbert series and some highest weight generating functions are computed.
This section also 
studies the one-dimensional members of these three families which are the Klein $A_3,\;E_6,$ and $E_8$ singularities, respectively.

In Section \ref{sec:CoulHS}, the computations of the Coulomb branch Hilbert series are presented. In the anomaly-free cases, the Coulomb branches are D-type Klein singularities; in the anomalous cases, the formal Hilbert series instead gives a putative $\mathbb Z_2$ cover at the level of Hilbert series.

In Section \ref{sec:conc}, some conclusions and future work are discussed.

In Appendix \ref{app_m1}, the non-normal variety $m_1$ is introduced. It appears as a transverse slice on the Higgs branch of the even $k$ cases.

In Appendix \ref{app_branching}, the decompositions of representations of $\mathrm{SU}(2)$ into irreducible representations of its finite subgroups are computed.


\section{Chain Polymerisation Construction of $g\sorm(N)$}
\label{sec:chain}
There are two different magnetic quivers for the $g\sorm(N)$ singularities depending on whether $N$ is even or odd.

The magnetic quiver for the $g\sorm(2n)$, where $N=2n$, also denoted by
$gd_n$, singularity is the following quiver 

\begin{equation}
    \begin{tikzpicture}
        
    \node[gauge, label=below:$2$] (2lres) at (0,-4){};
        \node[gauge, label=below:$2$] (2lmres) at (1,-4){};
        \node[gauge, label=below:$2$] (2lrres) at (2,-4){};
        \node[] (cdotsres) at (3,-4){$\cdots$};
        \node[gauge, label=below:$2$] (2rres) at (4,-4){};

        \draw[-] (2lres)--(2lmres)--(2lrres)--(cdotsres)--(2rres);
         \draw[transform canvas={yshift=2pt}] (2lres)--(2lmres);
        \draw[transform canvas={yshift=-2pt}] (2lres)--(2lmres);
        \draw[-] (0.5-0.1,-4-0.2)--(0.5+0.1,0-4)--(0.5-0.1,0.2-4);

        \node[] (Cdotsres) at (5,-4){$\cdots$};
        \node[gauge, label=below:$2$] (2Rres) at (6,-4){};
        \node[gauge, label=right:$1$] (1tRres) at ({6+cos(45)},{-4+sin(45)}){};
        \node[gauge, label=right:$1$] (1bRres) at ({6+cos(45)},{-4-sin(45)}){};

        \draw[-] (2rres)--(Cdotsres)--(2Rres)--(1tRres) (2Rres)--(1bRres);
         
         \draw [decorate, 
    decoration = {brace,
        raise=15pt,
        amplitude=5pt}] (2Rres) --  (2lres) node[pos=0.5,below=20pt,black]{$n-1$};
        
    \end{tikzpicture}
\end{equation}

whereas the magnetic quiver for the $g\sorm(2n+1)$, where $N=2n+1$, also 
denoted by
$gb_n$, singularity is the following quiver 

\begin{equation}
\begin{tikzpicture}
    \node[gauge, label=below:$2$] (2lres) at (0,-4){};
        \node[gauge, label=below:$2$] (2lmres) at (1,-4){};
        \node[gauge, label=below:$2$] (2lrres) at (2,-4){};
        \node[] (cdotsres) at (3,-4){$\cdots$};
        \node[gauge, label=below:$2$] (2rres) at (4,-4){};

        \draw[-] (2lres)--(2lmres)--(2lrres)--(cdotsres)--(2rres);
         \draw[transform canvas={yshift=2pt}] (2lres)--(2lmres);
        \draw[transform canvas={yshift=-2pt}] (2lres)--(2lmres);
        \draw[-] (0.5-0.1,-4-0.2)--(0.5+0.1,0-4)--(0.5-0.1,0.2-4);

        \node[] (Cdotsres) at (5,-4){$\cdots$};
        \node[gauge, label=below:$2$] (2Rres) at (6,-4){};
        \node[gauge, label=below:$1$] (1Rres) at (7,-4){};

        \draw[-] (2rres)--(Cdotsres)--(2Rres);
         \draw[transform canvas={yshift=1.3pt}] (2Rres)--(1Rres);
        \draw[transform canvas={yshift=-1.3pt}] (2Rres)--(1Rres);
        \draw[-] (6.5-0.1,-4-0.2)--(6.5+0.1,-4)--(6.5-0.1,-4+0.2);

         \draw [decorate, 
    decoration = {brace,
        raise=15pt,
        amplitude=5pt}] (1Rres) --  (2lres) node[pos=0.5,below=20pt,black]{$n+1$};
\end{tikzpicture}
\end{equation}

Both of these magnetic quivers give constructions for the $g\sorm(N)$ singularities as a moduli space of dressed monopole operators i.e. as a Coulomb branch of a $3d\;\mathcal N=4$ theory.

A natural hyper-Kähler quotient construction of magnetic quivers of this type involves a ``chain polymerisation'' of two different quivers \cite{Hanany:2024fqf}. The explicit constructions are summarised below.
\subsection{Chain Polymerisation Construction of $gd_{p+q}$}
First consider the finite $D_{q+2},\;q\geq 2$ quiver drawn below
\begin{equation}
\begin{tikzpicture}
        \node[gauge, label=below:$1$] (1l) at (0,0){};
        \node[gauge, label=below:$2$] (2l) at (1,0){};
        \node[] (cdots) at (2,0){$\cdots$};
        \node[gauge, label=below:$2$] (2r) at (3,0){};
        \node[gauge, label=right:$1$] (1tr) at ({3+cos(45)},{sin(45)}){};
        \node[gauge, label=right:$1$] (1br) at ({3+cos(45)},{-sin(45)}){};

        \draw[-] (1l)--(2l)--(cdots)--(2r)--(1tr) (2r)--(1br);

         \draw [decorate, 
    decoration = {brace,
        raise=15pt,
        amplitude=5pt}] (2r) --  (1l) node[pos=0.5,below=20pt,black]{$q$};
    \end{tikzpicture}
\end{equation}

The Coulomb branch is the moduli space of $2q$ free twisted hypermultiplets with $\sprm(2q)$ global symmetry.

Now consider the following quiver where $p\geq 1$
\begin{equation}
    \begin{tikzpicture}
        \node[gauge, label=below:$2$] (2l) at (0,0){};
        \node[gauge, label=below:$2$] (2lm) at (1,0){};
        \node[gauge, label=below:$2$] (2lr) at (2,0){};
        \node[] (cdots) at (3,0){$\cdots$};
        \node[gauge, label=below:$2$] (2r) at (4,0){};
        \node[gauge, label=below:$1$] (1r) at (5,0){};

        \draw[-] (2l)--(2lm)--(2lr)--(cdots)--(2r)--(1r);
         \draw[transform canvas={yshift=2pt}] (2l)--(2lm);
        \draw[transform canvas={yshift=-2pt}] (2l)--(2lm);
        \draw[-] (0.5-0.1,-0.2)--(0.5+0.1,0)--(0.5-0.1,0.2);

         \draw [decorate, 
    decoration = {brace,
        raise=15pt,
        amplitude=5pt}] (1r) --  (2l) node[pos=0.5,below=20pt,black]{$p+2$};
    \end{tikzpicture}\label{eq:H2pp2}
\end{equation}The Coulomb branch is, surprisingly, the moduli space of $2p+2$ free twisted hypermultiplets with $\sprm(2p+2)$ global symmetry which can be checked with the monopole formula \cite{Cremonesi:2013lqa}.

Both of these quivers have a tail of $(1)-(2)-\cdots$ and so may be chain polymerised \cite{Hanany:2024fqf} producing the $gd_{p+q}$ quiver, illustrated in \Figref{fig:gdChainPoly}.

\begin{figure}[h!]
    \centering
    \begin{tikzpicture}

    \node[gauge, label=below:$2$] (2l) at (0,0){};
        \node[gauge, label=below:$2$] (2lm) at (1,0){};
        \node[gauge, label=below:$2$] (2lr) at (2,0){};
        \node[] (cdots) at (3,0){$\cdots$};
        \node[gauge, label=below:$2$] (2r) at (4,0){};
        \node[gauge, label=below:$1$] (1r) at (5,0){};

        \draw[-] (2l)--(2lm)--(2lr)--(cdots)--(2r)--(1r);
         \draw[transform canvas={yshift=2pt}] (2l)--(2lm);
        \draw[transform canvas={yshift=-2pt}] (2l)--(2lm);
        \draw[-] (0.5-0.1,-0.2)--(0.5+0.1,0)--(0.5-0.1,0.2);

         \draw [decorate, 
    decoration = {brace,
        raise=15pt,
        amplitude=5pt}] (1r) --  (2l) node[pos=0.5,below=20pt,black]{$p+2$};

      \node[gauge, label=below:$1$] (1L) at (3,-2){};
        \node[gauge, label=below:$2$] (2L) at (4,-2){};
        \node[] (Cdots) at (5,-2){$\cdots$};
        \node[gauge, label=below:$2$] (2R) at (6,-2){};
        \node[gauge, label=right:$1$] (1tR) at ({6+cos(45)},{-2+sin(45)}){};
        \node[gauge, label=right:$1$] (1bR) at ({6+cos(45)},{-2-sin(45)}){};

        \draw[-] (1L)--(2L)--(Cdots)--(2R)--(1tR) (2R)--(1bR);

         \draw [decorate, 
    decoration = {brace,
        raise=15pt,
        amplitude=5pt}] (2R) --  (1L) node[pos=0.5,below=20pt,black]{$q$};

        \node (times) at (4,-1.25){$\times$};

        \node[gauge, label=below:$2$] (2lres) at (0,-4){};
        \node[gauge, label=below:$2$] (2lmres) at (1,-4){};
        \node[gauge, label=below:$2$] (2lrres) at (2,-4){};
        \node[] (cdotsres) at (3,-4){$\cdots$};
        \node[gauge, label=below:$2$] (2rres) at (4,-4){};

        \draw[-] (2lres)--(2lmres)--(2lrres)--(cdotsres)--(2rres);
         \draw[transform canvas={yshift=2pt}] (2lres)--(2lmres);
        \draw[transform canvas={yshift=-2pt}] (2lres)--(2lmres);
        \draw[-] (0.5-0.1,-4-0.2)--(0.5+0.1,0-4)--(0.5-0.1,0.2-4);

        \node[] (Cdotsres) at (5,-4){$\cdots$};
        \node[gauge, label=below:$2$] (2Rres) at (6,-4){};
        \node[gauge, label=right:$1$] (1tRres) at ({6+cos(45)},{-4+sin(45)}){};
        \node[gauge, label=right:$1$] (1bRres) at ({6+cos(45)},{-4-sin(45)}){};

        \draw[-] (2rres)--(Cdotsres)--(2Rres)--(1tRres) (2Rres)--(1bRres);
         
         \draw [decorate, 
    decoration = {brace,
        raise=15pt,
        amplitude=5pt}] (2Rres) --  (2lres) node[pos=0.5,below=20pt,black]{$p+q-1$};
        
    \end{tikzpicture}
    \caption{$\sprm(1)$ chain polymerisation of free fields to produce $gd_{p+q}$.}
    \label{fig:gdChainPoly}
\end{figure}

The action on the moduli spaces is a hyper-Kähler quotient by a diagonal $\sprm(1)\simeq\surm(2)$ symmetry. The construction is \begin{equation}
    \left(\mathbb H^{2q}\times\mathbb H^{2p+2}\right)/\!/\!/\sprm(1)=gd_{p+q}\label{eq:gdnHKQ}
\end{equation}

This may be verified through Hilbert series computations using the following embeddings to specify the $\sprm(1)$ action.\begin{align}
    [1,0,\cdots,0]_{\sprm(2q)}&\rightarrow [1,0,\cdots,0]_{\sorm(2q)}[1]_{\sprm(1)}\nonumber\\
    [1,0,\cdots,0]_{\sprm(2p+2)}&\rightarrow [1,0,\cdots,0]_{\sorm(2p)}[1]_{\sprm(1)}+[3]_{\sprm(1)}\label{eq:gdnEmbed}
\end{align}where the trivial representation is suppressed.

\subsection{Chain Polymerisation Construction of $gb_{p+q}$}
A very similar $\sprm(1)$ chain polymerisation construction exists for $gb_{p+q}$. It starts with the finite $B_{q+2},\;q\geq 1$ quiver drawn below \begin{equation}
    \begin{tikzpicture}
        \node[gauge, label=below:$1$] (1l) at (0,0){};
        \node[gauge, label=below:$2$] (2l) at (1,0){};
        \node[] (cdots) at (2,0){$\cdots$};
        \node[gauge, label=below:$2$] (2r) at (3,0){};
        \node[gauge, label=below:$1$] (1r) at (4,0){};

        \draw[-] (1l)--(2l)--(cdots)--(2r);
         \draw[transform canvas={yshift=1.3pt}] (2r)--(1r);
        \draw[transform canvas={yshift=-1.3pt}] (2r)--(1r);
        \draw[-] (3.5-0.1,-0.2)--(3.5+0.1,0)--(3.5-0.1,0.2);

         \draw [decorate, 
    decoration = {brace,
        raise=15pt,
        amplitude=5pt}] (1r) --  (1l) node[pos=0.5,below=20pt,black]{$q+2$};
    \end{tikzpicture}
\end{equation}

The Coulomb branch is the moduli space of $2q+1$ free twisted hypermultiplets with $\sprm(2q+1)$ global symmetry.

Then perform the chain polymerisation with the quiver in \eqref{eq:H2pp2} to construct the $gb_{p+q}$ quiver as shown in \Figref{fig:gbChainPoly}.

\begin{figure}[h!]
    \centering
    \begin{tikzpicture}

    \node[gauge, label=below:$2$] (2l) at (0,0){};
        \node[gauge, label=below:$2$] (2lm) at (1,0){};
        \node[gauge, label=below:$2$] (2lr) at (2,0){};
        \node[] (cdots) at (3,0){$\cdots$};
        \node[gauge, label=below:$2$] (2r) at (4,0){};
        \node[gauge, label=below:$1$] (1r) at (5,0){};

        \draw[-] (2l)--(2lm)--(2lr)--(cdots)--(2r)--(1r);
         \draw[transform canvas={yshift=2pt}] (2l)--(2lm);
        \draw[transform canvas={yshift=-2pt}] (2l)--(2lm);
        \draw[-] (0.5-0.1,-0.2)--(0.5+0.1,0)--(0.5-0.1,0.2);

         \draw [decorate, 
    decoration = {brace,
        raise=15pt,
        amplitude=5pt}] (1r) --  (2l) node[pos=0.5,below=20pt,black]{$p+2$};

     \node[gauge, label=below:$1$] (1L) at (3,-2){};
        \node[gauge, label=below:$2$] (2L) at (4,-2){};
        \node[] (Cdots) at (5,-2){$\cdots$};
        \node[gauge, label=below:$2$] (2R) at (6,-2){};
        \node[gauge, label=below:$1$] (1R) at (7,-2){};

        \draw[-] (1L)--(2L)--(Cdots)--(2R);
         \draw[transform canvas={yshift=1.3pt}] (2R)--(1R);
        \draw[transform canvas={yshift=-1.3pt}] (2R)--(1R);
        \draw[-] (6.5-0.1,-2-0.2)--(6.5+0.1,-2)--(6.5-0.1,-2+0.2);

         \draw [decorate, 
    decoration = {brace,
        raise=15pt,
        amplitude=5pt}] (1R) --  (1L) node[pos=0.5,below=20pt,black]{$q+2$};

        \node (times) at (4,-1.25){$\times$};

        \node[gauge, label=below:$2$] (2lres) at (0,-4){};
        \node[gauge, label=below:$2$] (2lmres) at (1,-4){};
        \node[gauge, label=below:$2$] (2lrres) at (2,-4){};
        \node[] (cdotsres) at (3,-4){$\cdots$};
        \node[gauge, label=below:$2$] (2rres) at (4,-4){};

        \draw[-] (2lres)--(2lmres)--(2lrres)--(cdotsres)--(2rres);
         \draw[transform canvas={yshift=2pt}] (2lres)--(2lmres);
        \draw[transform canvas={yshift=-2pt}] (2lres)--(2lmres);
        \draw[-] (0.5-0.1,-4-0.2)--(0.5+0.1,0-4)--(0.5-0.1,0.2-4);

        \node[] (Cdotsres) at (5,-4){$\cdots$};
        \node[gauge, label=below:$2$] (2Rres) at (6,-4){};
        \node[gauge, label=below:$1$] (1Rres) at (7,-4){};

        \draw[-] (2rres)--(Cdotsres)--(2Rres);
         \draw[transform canvas={yshift=1.3pt}] (2Rres)--(1Rres);
        \draw[transform canvas={yshift=-1.3pt}] (2Rres)--(1Rres);
        \draw[-] (6.5-0.1,-4-0.2)--(6.5+0.1,-4)--(6.5-0.1,-4+0.2);

         \draw [decorate, 
    decoration = {brace,
        raise=15pt,
        amplitude=5pt}] (1Rres) --  (2lres) node[pos=0.5,below=20pt,black]{$p+q+1$};
        
    \end{tikzpicture}
    \caption{$\sprm(1)$ chain polymerisation of free fields to produce $gb_{p+q}$.}
    \label{fig:gbChainPoly}
\end{figure}

The action on the moduli spaces is a hyper-Kähler quotient by a diagonal $\sprm(1)\simeq\surm(2)$ symmetry. The construction is \begin{equation}
    \left(\mathbb H^{2q+1}\times\mathbb H^{2p+2}\right)/\!/\!/\sprm(1)=gb_{p+q}\label{eq:gbnHKQ}
\end{equation}

This may be verified through Weyl integration computations on the Hilbert series using the following embeddings to specify the $\sprm(1)$ action.\begin{align}
    [1,0,\cdots,0]_{\sprm(2q+1)}&\rightarrow [1,0,\cdots,0]_{\sorm(2q+1)}[1]_{\sprm(1)}\nonumber\\
    [1,0,\cdots,0]_{\sprm(2p+2)}&\rightarrow [1,0,\cdots,0]_{\sorm(2p)}[1]_{\sprm(1)}+[3]_{\sprm(1)}\label{eq:gbnEmbed}
\end{align}where the trivial representation is suppressed.

\section{Electric Quiver for $g\sorm(N),\;h\sorm(N),$ and $i\sorm(N)$}
\label{sec:elecquiver}
The chain polymerisation constructions in the previous Section give an explicit hyper-Kähler quotient construction for the $g\sorm(N)$ singularities from free spaces in the forms of \eqref{eq:gdnHKQ} and \eqref{eq:gbnHKQ}, for even and odd $N$, respectively, together with the embeddings of $\sprm(1)$ into those free spaces in \eqref{eq:gdnEmbed} and \eqref{eq:gbnEmbed}, respectively.

These hyper-Kähler quotient constructions and the embeddings immediately provide an electric quiver, i.e. a Higgs branch hyper-Kähler quotient construction, for the $g\sorm(N)$ singularity
\begin{equation}
    \begin{tikzpicture}
        \node[gaugeb, label=below:$\sprm(1)$] (sp1) at (0,0){};
        \node[flavourr, label=above:$\sorm(N)$] (son) at (0,1.5){};
        \node[flavourr, label=right:$\orm(1)$] (o1) at (1.5,0){};
        \draw[-] (sp1)--(son);
        \draw[-,color=red] (sp1)--(o1)node[midway, above, color=black]{$\mathrm{Sym}^3$};
    \end{tikzpicture}
\end{equation}This quiver is ADHM-like in the sense that there are two matter representations; fundamental and $\mathrm{Sym}^3$. However the interpretation of $\mathrm{Sym}^3$ matter is unclear in a brane system. Importantly the red line refers to a half-hypermultiplet which makes it clear that the electric quiver suffers a Witten anomaly \cite{Witten:1982fp} for odd $N$ since the Dynkin index for the $[3]$ of $\sprm(1)$ is $T\left([3]\right)=5$ which is an integer. Nevertheless, the resulting moduli space is an isolated symplectic singularity, which may appear in the Hasse diagram of non-anomalous theories, the moduli space itself is still interesting from a mathematical point of view.

Despite the anomaly for odd $N$, take the view that the quiver is a tool to compute Hilbert series of moduli spaces, giving only physical interpretation when $N$ is even.

Taking this approach one finds agreement that the Higgs branch of the electric quiver is the $g\sorm(N)$ singularity via computation of Hilbert series and HWG.

At this point, the data defining the $g\sorm(N)$ singularities are equivalently contained in the magnetic and electric quivers. However, in \cite{Bourget:2025wsp} it was argued that there are no further unitary magnetic quivers which give rise to isolated conical symplectic singularities. There is no such constraint for the electric quiver and some natural generalisations may result in new isolated conical symplectic singularities for which Coulomb branch constructions are not known. 

The electric quiver above can be generalised by considering $\mathrm{Sym}^k$ matter for all integers $k$. The trivial cases of $k=1,2$ are $\sprm(1)$ SQCD and ADHM, respectively, however for $k\geq 3$ these theories have no particular name. The quiver is drawn below \begin{equation}
    \begin{tikzpicture}
        \node[gaugeb, label=below:$\sprm(1)$] (sp1) at (0,0){};
        \node[flavourr, label=above:$\sorm(N)$] (son) at (0,1.5){};
        \node[flavourr, label=right:$F$] (o1) at (1.5,0){};
        \draw[-] (sp1)--(son);
        \draw[-,color=red] (sp1)--(o1)node[midway, above, color=black]{$\mathrm{Sym}^k$};
    \end{tikzpicture}\label{quiv:SymkFamily}
\end{equation}where the flavour symmetry group $F$ is $\orm(1)$ if $k$ is odd and $\sprm(1)$ if $k$ is even. This is because the representation $\mathrm{Sym}^k$ is pseudoreal for odd $k$ and real for even $k$. The theories with odd $k$ may also suffer a Witten anomaly, which does not affect construction of moduli spaces. 

The Higgs branch dimension is $N+(k+1)/2-3$ for the case of odd $k$ and $N+(k+1)-3$ for even $k$.

As it turns out only the cases with $k=1,3,5,7$ correspond to isolated conical symplectic singularities (ICSS). With $k=1,3$ these are minimal nilpotent orbit closures and the $g\sorm(N)$ singularities, respectively. The cases of $k=5,7$ are new ICSS which shall be referred to as $h\sorm(N)$ and $i\sorm(N)$, respectively. This claim is verified in the next section.

For the case of odd $k$ it is also possible to gauge the $\orm(1)$ flavour symmetry factor. This produces the quiver below whose Higgs branch is the $\orm(1)\simeq\mathbb Z_2$ quotient of the $g\sorm(N), \;h\sorm(N),$ and $i\sorm(N)$ singularities. These are generically not isolated symplectic singularities.

\begin{equation}
    \begin{tikzpicture}
        \node[gaugeb, label=below:$\sprm(1)$] (sp1) at (0,0){};
        \node[flavourr, label=above:$\sorm(N)$] (son) at (0,1.5){};
        \node[gauger, label=below:$\orm(1)$] (o1) at (1.5,0){};
        \draw[-] (sp1)--(son);
        \draw[-,color=red] (sp1)--(o1)node[midway, above, color=black]{$\mathrm{Sym}^{k}$};
    \end{tikzpicture}\label{quiv:SymkZ2Family}
\end{equation}

\subsection{Higgs Mechanism}
The claim that the magnetic quivers in \Figref{fig:gdChainPoly} and \Figref{fig:gbChainPoly} correspond to isolated conical symplectic singularities was verified through the decay and fission algorithm \cite{Bourget:2023dkj,Bourget:2024mgn} in \cite{Bourget:2025wsp}.

The corresponding check for the electric quivers in \Quiver{quiv:SymkFamily} is through the Higgs mechanism. If the Higgs branch corresponds to an isolated symplectic singularity, then the gauge group $\sprm(1)$ must only break to the trivial group. If the Higgs branch is not an isolated symplectic singularity, then $\sprm(1)$ may break either to a continuous subgroup $\urm(1)\rtimes\mathbb Z_2\cong \mathrm{Pin}(2)$ or $\urm(1)$, or to a discrete subgroup $\Gamma_{ADE}$. The strategy is to first attempt to break the $\sprm(1)$ gauge group to $\urm(1)$, and then, if necessary, to a discrete subgroup. The breaking pattern is analysed separately for odd and even $k$.

\subsubsection{Odd $k$}
The decompositions of irreducible representations from $\sprm(1)$ to $\mathrm{Pin}(2)$ used here are
\begin{align}
    [1]_{\sprm(1)}&\rightarrow \rho_1\nonumber\\
    [k]_{\sprm(1)}&\rightarrow \rho_k+\rho_{k-2}+\cdots+\rho_1\nonumber\\
    [2]_{\sprm(1)}&\rightarrow 1+\underbrace{\rho_2}_{\text{Acquire mass}}\label{SU2toPin2mass}
\end{align}
where $\rho_j$ is a two-dimensional irreducible representation on which $\urm(1)$ acts with charge $j$, and $\mathbb{Z}_2$ acts by the Weyl reflection. From this decomposition, it is immediately clear that $\sprm(1)$ cannot be broken to $\mathrm{Pin}(2)$. Indeed, the W-bosons cannot acquire mass because of the incompatible irreducible representations: the matter fields contain only $\rho_{\mathrm{odd}}$, whereas the W-bosons transform in $\rho_2$.

The further decompositions of irreducible representations from $\mathrm{Pin}(2)$ to $\urm(1)$ are
\begin{equation}
    \rho_j\to q^j +q^{-j}
\end{equation}
where $q$ is a $\urm(1)$ fugacity.
In total, the decompositions from $\sprm(1)$ to $\urm(1)$ are
\begin{align}
    [1]_{\sprm(1)}&\rightarrow q+q^{-1}\nonumber\\
    [k]_{\sprm(1)}&\rightarrow q^k+q^{k-2}+\cdots+q+q^{-1}+\cdots q^{2-k}+q^{-k}\nonumber\\
    [2]_{\sprm(1)}&\rightarrow 1+\underbrace{q^2+q^{-2}}_{\text{Acquire mass}}\label{SU2toU1mass}
\end{align}
where the W-bosons must acquire mass in the Higgs mechanism. Again, it is immediately clear that $\sprm(1)$ cannot be broken to $\urm(1)$, since the W-bosons cannot acquire mass because of the incompatible charge assignments: the matter fields have odd charges, whereas the W-bosons have charges $\pm2$.

Therefore, the $\sprm(1)$ gauge symmetry can only be broken to a discrete subgroup, namely one of the $\Gamma_{ADE}$ discrete groups.

Let us first discuss the $E$-type subgroups, which are the binary tetrahedral ($2\mathrm T$), binary octohedral ($2\mathrm O$), and binary icosahedral ($2\mathrm I$) groups corresponding to $E_{6,7,8}$, respectively. The branching rule for $[k]_{\sprm(1)}$ into irreducible representations of  $2\mathrm T$, $2\mathrm O$, and $2\mathrm I$ is discussed in Appendix \ref{app_branching} where it is shown that the decomposition of $[k]_{\sprm(1)}$ for odd $k$ contains a different set of irreducible representations from the decomposition for even $k$. Hence the components of $[2]_{\sprm(1)}$ cannot acquire mass when $k$ is odd, and $\sprm(1)$ cannot be broken to $2\mathrm T$, $2\mathrm O$, or $2\mathrm I$.

Now consider the $D$-type subgroups. Since $\mathrm{Dic}_l$ is also a subgroup of $\mathrm{Pin}(2)$, it is convenient to work with the decomposition of irreducible representations from $\mathrm{Pin}(2)$ to $\mathrm{Dic}_l$, which is
\begin{align}
    \rho_j&\rightarrow 1+\chi_1,&\ j\ \mathrm{mod}\ 2l=0\\
    \rho_j&\rightarrow \chi_2+\chi_3,&\ j\ \mathrm{mod}\ 2l=l\\
    \rho_j&\rightarrow \rho_{r},&\ j\ \mathrm{mod}\ 2l =r <l\\
    \rho_j&\rightarrow \rho_{2l-r},&\ j\ \mathrm{mod}\ 2l=r >l
\end{align}
where $\chi_i$ are one-dimensional irreducible representations, and $\rho_r$ is the natural restriction of $\rho_j$. It is clear that the $\rho_2$ irreducible representation of $\mathrm{Pin}(2)$ cannot acquire mass by absorbing components of $\rho_{\mathrm{odd}}$ after decomposing to $\mathrm{Dic}_l$. Hence $\sprm(1)$ cannot be broken to $\mathrm{Dic}_l$.


The only remaining choice is the $A$-type subgroup. Take $\mathbb Z_l\subset\mathrm U(1)$, where $l$ is an odd integer. The branching follows the same pattern as above, where $q=e^{2\pi i/l}$ is now a primitive $l$-th root of unity, and the charge $i$ representation is identified with the charge $i\ \mathrm{mod}\ l$ representation of $\mathbb Z_l$. Since there is no adjoint representation of $\mathbb Z_l$, all components of the W-bosons acquire mass:
\begin{equation}
    [2]_{\sprm(1)}\rightarrow \underbrace{1+q^2+q^{-2}}_{\text{Acquire mass}}
\end{equation}
Note that there are two copies of the real $[2]_{\sprm(1)}$ representation. The charge $0$ components of the W-bosons can acquire mass by eating charge $\pm l$ matter. The charge $\pm2$ components of the W-bosons can acquire mass by eating charge $\pm l\pm2$ matter.
For this breaking pattern to be allowed, another condition must be satisfied: the transition must be non-trivial. This non-triviality condition is equivalent to the existence of vevs enabling the breaking pattern. It imposes the constraint that the components of the matter fields transforming trivially under such a $\mathbb Z_l$ have complex dimension greater than $6$, so that matter fields remain after the W-bosons acquire mass.

No such $\mathbb Z_l$ can be found for $k=3,5,7$. For $k=9$, the gauge group can be broken to a $\mathbb Z_3$ subgroup. For $k\geq9$, a $\mathbb Z_l$ subgroup can be broken to if and only if $l=\mathrm{gcd}(k_1,k_2)\neq1$ for some pair $k_1,k_2\in \mathbf{k}:=\{1,3,\dots,k\}$. This Higgsing is, in general, not minimal.
To find the minimal Higgsing, take $\mathbf{g}$ to be the set of all distinct non-trivial values of $\mathrm{gcd}(k_1,k_2)$. It has a subset $\mathbf{l}\subset\mathbf{g}$ such that, for all $l_1,l_2 \in\mathbf{l}$, one has $l_1 \nmid l_2$ and $l_2 \nmid l_1$. This subset $\mathbf{l}$ is the set of all allowed $l$ for minimal Higgsing.
The residual theory is expressed by the quiver
 \begin{equation}
    \begin{tikzpicture}
        \node[gauge, label=left:$\mathbb{Z}_l$] (u1) at (0,0){};
        \node[flavour, label=above:$F_1$] (n) at (0,1.5){};
        \node[flavour, label=right:$F_3$] (2t) at ({1.5*cos(30)},{1.5*sin(30)}){};
        \node (cdots) at ({1.5*cos(30)},0){$\vdots$};
        \node[flavour, label=right:$F_{l-2}$] (2b) at ({1.5*cos(30)},{1.5*-sin(30)}){};

        \draw[-] (u1)--(n)node[midway, left]{$1$} (u1)--(2t)node[midway, above]{$3$} (u1)--(2b)node[midway, below]{$l\!-\!2$};
    \end{tikzpicture}
\end{equation}
where the labels next to the edges refer to the charge of the bifundamental under $\mathbb{Z}_l$. Take $k=ml+r$, where $0\leq r<l$. Then the flavour multiplicities are
\begin{equation}
    F_i=m+\delta_{i,1}-\delta_{i,l-2}-\delta_{i,2}+\sum_{j=1}^r\delta_{i,j}.
\end{equation}
In general, the Higgs branch of the residual theory is not an isolated singularity; its Higgsing and 3d mirror are studied in \cite{Grimminger:2025fgj}.

It is interesting to see what the transverse theory looks like. The normalised transverse theory\footnote{More precisely, the action of the original gauge group on the transverse matter is not always equivalent to the action of the normaliser subgroup here. When they mismatch, the transverse slice is a non-normal variety and there is no well-defined gauge theory description. For instance, the non-normal slice $m_1$ appears for even $k$ below, and its generalisation to the $m_g$ slices appears in \cite{Bennett:2024loi}.} has gauge group
\begin{equation}
    N_{\mathrm{\surm(2)}}(\mathbb{Z}_l)/\mathbb{Z}_l
    \cong \left(\mathrm{U}(1)\rtimes\mathcal{W}_{\mathrm{SU}(2)}\right)/\mathbb{Z}_l
    \cong \mathrm{Pin}(2)/\mathbb{Z}_l .
\end{equation}
The matter fields are the components which transform trivially under $\mathbb{Z}_l$. Note that the Weyl $\mathbb{Z}_2$ action depends on whether $k\ \mathrm{mod}\ 4=1$ or $3$, and it does not commute with the $\mathrm U(1)$ action.
The transverse theory is expressed by the quiver
 \begin{equation}
    \begin{tikzpicture}
        \node[gauge, label=left:$\mathrm{Pin}(2)/\mathbb{Z}_l$] (u1) at (0,0){};
        \node[flavour, label=above:$m$] (n) at (0,1.5){};
        \node[flavour, label=right:$m$] (2t) at ({1.5*cos(30)},{1.5*sin(30)}){};
        \node (cdots) at ({1.5*cos(30)},0){$\vdots$};
        \node[flavour, label=right:$m$] (2b) at ({1.5*cos(30)},{1.5*-sin(30)}){};

        \draw[-] (u1)--(n)node[midway, left]{$\rho_l$} (u1)--(2t)node[midway, above]{$\rho_{2l}$} (u1)--(2b)node[midway, below]{$\rho_{ml}$};
    \end{tikzpicture}
\end{equation}
The Higgs branch of the transverse theory is an isolated singularity if $l\in\mathbf{l}$, i.e. if it is obtained by minimal Higgsing.


\subsubsection{Even $k$ with $k=0\ \mathrm{mod}\ 4$}
Taking the quiver in \eqref{quiv:SymkFamily} with even $k$ produces different results. The branching rules, which is clear from the following discussion, suggest splitting the analysis into two cases: $k=0\ \mathrm{mod}\ 4$ and $k=2\ \mathrm{mod}\ 4$. In this section we discuss the case $k=0\ \mathrm{mod}\ 4$.

The decompositions of irreducible representations from $\sprm(1)$ to $\mathrm{Pin}(2)$ used here are the same as those in \eqref{SU2toPin2mass}, except that $k$ is even here with $k=0\ \mathrm{mod}\ 4$
\begin{align}
    [1]_{\sprm(1)}&\rightarrow \rho_1\nonumber\\
    [k]_{\sprm(1)}&\rightarrow \rho_k+\rho_{k-2}+\cdots+\rho_2+1\nonumber\\
    [2]_{\sprm(1)}&\rightarrow 1+\underbrace{\rho_2}_{\text{Acquire mass}}
\end{align}
Note that there are two copies of the real $[k]$ representation. Performing the Higgs mechanism to obtain the resulting gauge theory gives
\begin{equation}
    N[1]_{\sprm(1)}+2[k]-2(\rho_2)=N(\rho_1)+2(\rho_k+\rho_{k-2}+\cdots+\rho_4)+2
\label{eq_evenmassiveeatupPin2}
\end{equation}
where the degrees of freedom are counted as complex rather than quaternionic.
The two complex singlets on the right-hand side indicate that the resulting gauge theory lies on a symplectic leaf of quaternionic dimension one.
This immediately suggests that even $k$ does not correspond to an isolated symplectic singularity. This is expected since the flavour symmetry of the original theory is $\sprm(1)\times\sorm(N)$, and isolated singularities are not expected to have a global symmetry which is a product of non-Abelian factors.

The residual theory is expressed by the quiver
 \begin{equation}
    \begin{tikzpicture}
        \node[gauge, label=left:$\mathrm{Pin}(2)$] (u1) at (0,0){};
        \node[flavour, label=above:$1$] (n) at (0,1.5){};
        \node[flavour, label=right:$1$] (2t) at ({1.5*cos(30)},{1.5*sin(30)}){};
        \node (cdots) at ({1.5*cos(30)},0){$\vdots$};
        \node[flavour, label=right:$1$] (2b) at ({1.5*cos(30)},{1.5*-sin(30)}){};
        \draw[-] (u1)--(n)node[midway, left]{$\rho_1$} (u1)--(2t)node[midway, above]{$\rho_{4}$} (u1)--(2b)node[midway, below]{$\rho_{k}$};
    \end{tikzpicture}
\end{equation}
\sloppy The transverse theory is more subtle here. The normalised theory has gauge group
\begin{equation}
    N_{\mathrm{\surm(2)}}(\mathrm{U}(1))/\mathrm{Pin}(2)\cong \{1\}.
\end{equation}
This theory is simply a free hypermultiplet, whose Higgs branch is $\mathbb C^2$.
However, this Higgs branch only determines the normalisation of the transverse slice. The actual transverse slice is a non-normal variety $m_1$, which can be computed from the coordinate ring; see Appendix \ref{app_m1} for details.

The further Higgsing from $\mathrm{Pin(2)}$ to $\urm(1)$ is not allowed, since no new trivial component appears after decomposition.

Higgsing to finite $\Gamma_{ADE}$ subgroups is, in general, allowed for sufficiently large $k$. More precisely, after restricting $[k]_{\sprm(1)}$ to a subgroup $\Gamma$, the W-boson representation $[2]_{\sprm(1)}|_{\Gamma}$ can acquire mass whenever its non-trivial irreducible components occur in the matter sector $[k]_{\sprm(1)}|_{\Gamma}$ with sufficient multiplicity, and the invariant components provide the required non-trivial vevs. For $D$-type subgroups this criterion is read from the coefficients $a_{n,r}^\Gamma(k)$ in Appendix \ref{app_branching}; for $E_6,E_7,E_8$ it is read from the generating functions $P^{\Gamma}_{n,r}(t)$. Since $k$ is even, only the even-parity irreducible representations of Appendix \ref{app_branching} appear, matching the parity sector of $[2]_{\sprm(1)}$.

\subsubsection{Even $k$ with $k=2\ \mathrm{mod}\ 4$}
In this section we discuss the case $k=2\ \mathrm{mod}\ 4$.

The decompositions of irreducible representations from $\sprm(1)$ to $\mathrm{Pin}(2)$ used here are
\begin{align}
    [1]_{\sprm(1)}&\rightarrow \rho_1\\
    [k]_{\sprm(1)}&\rightarrow \rho_k+\rho_{k-2}+\cdots+\rho_2+\epsilon\\
    [2]_{\sprm(1)}&\rightarrow 1+\underbrace{\rho_2}_{\text{Acquire mass}}
\end{align}
where $\epsilon$ is a one-dimensional irreducible representation on which $\urm(1)$ acts trivially and $\mathbb Z_2$ acts by a sign.
As in \eqref{eq_evenmassiveeatupPin2}, the $\rho_2$ component from $[2]_{\sprm(1)}$ can absorb the $\rho_2$ component from $[k]_{\sprm(1)}$ to acquire mass.
However, there are no singlets on the right-hand side. This means that there are no vevs to enable this breaking pattern, and hence $\sprm(1)$ cannot be Higgsed to $\mathrm{Pin}(2)$.

The further decomposition of irreducible representations from $\mathrm{Pin}(2)$ to $\urm(1)$ is
\begin{align}
    [1]_{\sprm(1)}&\rightarrow q+q^{-1}\\
    [k]_{\sprm(1)}&\rightarrow q^k+q^{k-2}+\cdots+q^2+1+q^{-2}+\cdots +q^{2-k}+q^{-k}\\
    [2]_{\sprm(1)}&\rightarrow 1+\underbrace{q^2+q^{-2}}_{\text{Acquire mass}}
\end{align}
Note that there are two copies of the real $[k]$ representation. Performing the Higgs mechanism to obtain the resulting gauge theory gives
\begin{equation}
    n[1]_{\sprm(1)}+2[k]-2(q^2+q^{-2})=n(q+q^{-1})+2(q^{k}+q^{k-2}+\cdots+q^4+\textrm{Reciprocal})+2
\label{eq_evenmassiveeatup}
\end{equation}
where the degrees of freedom are counted as complex rather than quaternionic.
The two complex singlets on the right-hand side indicate that the resulting gauge theory lies on a symplectic leaf of quaternionic dimension one. This immediately suggests that even $k$ does not correspond to an isolated symplectic singularity. This is expected since the flavour symmetry of the original theory is $\sprm(1)\times\sorm(n)$, and isolated singularities are not expected to have a global symmetry which is a product of non-Abelian factors.

The normalised theory has gauge group
\begin{equation}
    N_{\mathrm{\surm(2)}}(\mathrm{U}(1))/\mathrm{U}(1)\cong \mathcal{W}_{\mathrm{SU}(2)}  \cong \mathbb{Z}_2 .
\end{equation}
The matter fields are the components of weight $0$. Its Higgs branch is $A_1\cong \mathbb C^2/\mathbb Z_2$.

Higgsing to finite $\Gamma_{ADE}$ subgroups is, in general, allowed. The precise condition is again determined by the branching of $[k]_{\sprm(1)}$ to $\Gamma$: the irreducible components of $[2]_{\sprm(1)}|_{\Gamma}$ must be present in $[k]_{\sprm(1)}|_{\Gamma}$ so that the W-bosons can acquire mass, and the invariant sector must be large enough to support a non-trivial Higgs vev. For each D- or E-type subgroup, this criterion can be read from the multiplicities $a_{n,r}^\Gamma(k)$, equivalently from the coefficients of the generating functions $P_{n,r}^\Gamma(t)$ computed in Appendix \ref{app_branching}. Since $k$ is even, the relevant irreducible representations lie in the same central-parity sector as those appearing in $[2]_{\sprm(1)}|_{\Gamma}$.

\section{Higgs Branch Hilbert Series}
\label{sec:HiggsBranchHS}
In this Section, Hilbert series and HWGs for some cases of the $g\sorm(N),\;h\sorm(N),$ and $i\sorm(N)$ singularities are presented. These are computed using Weyl integration from the electric quiver \Quiver{quiv:SymkFamily} with the appropriate values of $k$ and $N$.
\subsection{$g\sorm(N)$}
This is the case with $k=3$, the Hilbert series and HWG match those of the magnetic quiver \cite{Bourget:2025wsp} for all $N\geq3$ which is the regime of validity for the magnetic quiver. However, the magnetic quiver is not able to capture the limiting one dimensional singularity in the $g\sorm(N)$ family which should be a Klein singularity when $N=2$. This is not a problem with the electric quiver \Quiver{quiv:SymkFamily}.

In this case the Hilbert series is computed as \begin{equation}
    \hs\left[g\sorm(2)\right]= \frac{1 - t^8}{(1 - t^2) (1 - qt^4) (1 - t^4/q)}
\end{equation}where $q$ is a $\urm(1)$ fugacity. The Hilbert series identifies the moduli space as the Klein $A_3\simeq D_3$ singularity.

For $N \geq 3$, the electric quiver we propose produces concordant Hilbert series with the magnetic computation in \emph{op.~cit.}, namely:
\begin{equation}
    \hwg\left[g\sorm(N)\right]= \begin{cases}\pe\left[\mu^2 t^2 + (1+\mu^6)t^4 + \mu^6 t^6 - \mu^{12} t^{12}\right],&N=3\\ \pe\left[(\mu_1^2 + \mu_2^2) t^2 + (1 + \mu_1^3 \mu_2^3) t^4 + \mu_1^3 \mu_2^3 t^6 - 
 \mu_1^6 \mu_2^6 t^{12}\right],&N=4\\\pe\left[\mu_2^2t^2+(1+\mu_1^2+\mu_1^3)t^4+\mu_1^3t^6-\mu_1^6t^{12}\right],&N=5\\\pe\left[\mu_2\mu_3t^2+(1+\mu_1^2+\mu_1^3)t^4+\mu_1^3t^6-\mu_1^6t^{12}\right],&N=6\\\pe\left[\mu_2t^2+(1+\mu_1^2+\mu_1^3)t^4+\mu_1^3t^6-\mu_1^6t^{12}\right],&N\geq 7\end{cases}
\end{equation}
\subsection{$h\sorm(N)$}
The $h\sorm(N)$ singularity is the case with $k=5$ in \Quiver{quiv:SymkFamily}. There is a one dimensional member of this family when $N=1$. The Hilbert series is computed as\begin{equation}
    \hs\left[h\sorm(1)\right]=\frac{1-t^{24}}{(1-t^6)(1-t^8)(1-t^{12})}
\end{equation} which identifies the moduli space as the Klein $E_6$ singularity.

This is suggestive of the following three equivalent quotient constructions of this moduli space, one as a discrete quotient and two as hyper-Kähler quotients\begin{equation}\mathbb C^2/\Gamma_{E_6}=\mathbb C^{48}/\!/\!/\left[\urm(1)^2\times\urm(2)^3\times\urm(3)\right]=\mathbb C^{8}/\!/\!/\sprm(1)\end{equation}

The first is the usual discrete group quotient of $\mathbb C^2$ by the binary tetrahedral group. The second is the hyper-Kähler quotient construction by Kronheimer as the Higgs branch of the $\widehat E_6$ quiver. The third is the implication from the electric quiver introduced in this paper as the Higgs branch of an $\sprm(1)$ gauge theory.

In general for $N\geq1$, the unrefined Hilbert series takes the following form \begin{equation}
    \hs\left[h\sorm(N)\right]=\frac{P_{8N+4}(t)}{(1-t^4)^{N-1}(1-t^6)^N(1-t^8)}
\end{equation}where $P_{8N+4}(t)$ is a palindromial (that is, palindromic polynomial) of degree $8N+4$. Some examples of $P_{8N+4}(t)$ are given in Table \ref{tab:P8n}.

\begin{table}[h!]
    \centering
    \begin{tabular}{cc}
    \toprule
        $N$ & $P_{8N+4}(t)$  \\
        \midrule
        $1$ & $1+t^{12}$\\\midrule
        $2$ & $1 + t^2 + 4 t^6 + 10 t^8 + 6 t^{10} + 10 t^{12} + 4 t^{14} + t^{18} + t^{20}$\\\midrule
         $3$& $1 + 3 t^2 + 4 t^4 + 22 t^6 + 70 t^8 + 91 t^{10} + 122 t^{12} + 144 t^{14} + \cdots + t^{28}$\\\midrule
         $4$ & $\begin{aligned}1 &+ 6 t^2 + 17 t^4 + 84 t^6 + 323 t^8 + 677 t^{10} + 1193 t^{12} \\&+ 
 1924 t^{14} + 2509 t^{16} + 2650 t^{18} + \cdots + t^{36}\end{aligned}$\\\midrule
   $5$ &  $\begin{aligned}1 &+ 10 t^2 + 46 t^4 + 256 t^6 + 1175 t^8 + 3422 t^{10} + 7961 t^{12} + 
 16241 t^{14} \\&+ 27812 t^{16} + 39586 t^{18} + 49122 t^{20} + 53042 t^{22} + 
 \cdots + t^{44}\end{aligned}$\\\bottomrule
    \end{tabular}
    \caption{Numerator of unrefined Hilbert series of $h\sorm(N)$, $P_{8N+4}(t)$, for $N=1,2,3,4,5$.}
    \label{tab:P8n}
\end{table}

The HWG is not a PE of a polynomial, but it can be expressed as a product of a polynomial with a PE of a polynomial. \begin{equation}\hwg\left[h\sorm(N)\right]=f_N(t,\mu_1,\mu_2)\times\mathrm{PE}\left[g_N(t,\mu_1,\mu_2,\mu_3)\right]
\end{equation}where the polynomial $f_N(t,\mu_1,\mu_2)$ is a degree $34$ polynomial for all $N$ and $\mu_{1,2}$ are highest weight fugacities for $\sorm(N)$. In general, the functions $f_N$ are functions of only $t,\mu_1$ and likewise the functions $g_N$ are only functions of $t,\mu_1,\mu_2$. Dependence on other highest weight fugacities occurs for only low rank cases. For this reason consider the following function $f_N(t,\mu_1)$ for $N\geq 5$,

\begin{align}
    f_{N\geq 5}(t,\mu_1)&=1 + (\mu_1 +\mu_1^3) t^6 + (\mu_1 +\mu_1^2 +\mu_1^3 +\mu_1^4 +\mu_1^5) t^8 + (\mu_1^2 + 
   \mu_1^4 +\mu_1^6) t^{10} \nonumber\\&+ (\mu_1 +\mu_1^2 +\mu_1^3 +\mu_1^4 +\mu_1^5 +\mu_1^7 + 
   \mu_1^9) t^{12} + (\mu_1 +\mu_1^2 +\mu_1^3 +\mu_1^4 +\mu_1^5) t^{14} \nonumber\\&- (\mu_1^7 + 
   \mu_1^9 +\mu_1^{11}) t^{16} + (1 +\mu_1^2 +\mu_1^4) t^{18} - (\mu_1^6 +\mu_1^7 + 
   \mu_1^8 +\mu_1^9 +\mu_1^{10}) t^{20} \nonumber\\&- (\mu_1^2 +\mu_1^4 +\mu_1^6 +\mu_1^7 +\mu_1^8 + 
   \mu_1^9 +\mu_1^{10}) t^{22} - (\mu_1^5 +\mu_1^7 +\mu_1^9) t^{24} \nonumber\\&- (\mu_1^6 +\mu_1^7 + 
   \mu_1^8 +\mu_1^9 +\mu_1^{10}) t^{26} - (\mu_1^8 +\mu_1^{10}) t^{28} -\mu_1^{11} t^{34}
\end{align}
the above formula also applies for the case of $N=3, 4$ with the replacement of $\mu_1\leftrightarrow \mu^2, \mu_1\mu_2$, respectively.

Similarly the functions $g_N(t,\mu_1,\mu_2)$ for $N\geq 7$ are \begin{equation}
g_{N\geq 7}(t,\mu_1,\mu_2)=\mu_2 t^2 +(1+\mu_1^2) t^4+ \mu_1^5 t^6+ (1+\mu_1^6)t^8 +t^{12}
\end{equation}where for the cases of $N=5, 6$ one can replace $\mu_2\leftrightarrow \mu_2^2, \mu_2\mu_3$, respectively. However for the case of $N=3, 4$ we have 
\begin{align}
    g_3(t,\mu)&=\mu^2 t^2+ t^4+\mu^{{10}} t^6+  (1+\mu^{12})t^8 +t^{12}\\
    g_4(t,\mu_1,\mu_2)&=(\mu_1^2+\mu_2^2) t^2 +t^4+ \mu_1^5\mu_2^5 t^6+ (1+\mu_1^6\mu_2^6)t^8 +t^{12}
\end{align}

\subsection{$i\sorm(N)$}
The $i\sorm(N)$ singularity is the case with $k=7$ in \Quiver{quiv:SymkFamily}. Much like with the $g\sorm(N)$ and $h\sorm(N)$ singularities, there is also a member of the $i\sorm(N)$ singularity which corresponds to a Klein singularity, the case of $N=0$. The Hilbert series is
\begin{equation}
    \hs\left[i\sorm(0)\right]=\frac{1-t^{60}}{(1-t^{12})(1-t^{20})(1-t^{30})}
\end{equation}which identifies the moduli space as the Klein $E_8$ singularity. 

Just like the Klein $E_6$ singularity was found as a Higgs branch so too is the Klein $E_8$ singularity. There is also the analogous equivalence between moduli space constructions \begin{equation}\mathbb C^2/\Gamma_{E_8}=\mathbb C^{240}/\!/\!/\left[\urm(2)^2\times\urm(3)^2\times\urm(4)^2\times\urm(5)\times\urm(6)\right]=\mathbb C^{8}/\!/\!/\sprm(1)\end{equation}where the first is the usual discrete group quotient, the second is the construction of Kronheimer, and the third is a hyper-Kähler quotient with the action of $\sprm(1)$ specified by the matter representations in the electric quiver \Quiver{quiv:SymkFamily}.

In general for $N\geq 0$, the unrefined Hilbert series takes the following form \begin{equation}
    \hs\left[i\sorm(N)\right]=\frac{P_{12N+30}(t)}{(1-t^6)^{N}(1-t^8)^N(1-t^{12})(1-t^{20})}
\end{equation}where $P_{12N+30}(t)$ is a palindromial of degree $12N+30$. Some examples of $P_{12N+30}(t)$ are given in Table \ref{tab:P12N+30}.

\begin{table}[h!]
    \centering
    \begin{tabular}{cc}
    \toprule
        $N$ & $P_{12N+30}(t)$  \\
        \midrule
        $0$ & $1+t^{30}$\\\midrule
        $1$ & $\begin{aligned}&1 + 4 t^8 + t^{10} + 8 t^{12} + 5 t^{14} + 8 t^{16} + 9 t^{18} + 8 t^{20} \\&+ 
 8 t^{22} + 9 t^{24} + 8 t^{26} + 5 t^{28} + 8 t^{30} + t^{32} + 4 t^{34} + t^{42}\end{aligned}$\\\midrule
         $2$& $\begin{aligned}1 &+ t^2 + t^4 + 3 t^6 + 26 t^8 + 32 t^{10} + 88 t^{12} + 108 t^{14} + 
 173 t^{16} \\&+ 220 t^{18} + 284 t^{20} + 305 t^{22} + 360 t^{24} + 364 t^{26} + \cdots +t^{54}\end{aligned}$\\\midrule
         $3$ & $\begin{aligned}1 &+ 3 t^2 + 6 t^4 + 17 t^6 + 115 t^8 + 258 t^{10} + 686 t^{12} + 1270 t^{14} + 2331 t^{16} + 3721 t^{18} + 5706 t^{20} \\&+ 7653 t^{22} + 10065 t^{24} + 12003 t^{26} + 13890 t^{28} + 15138 t^{30} + 15771 t^{32} + \cdots + t^{66}\end{aligned}$\\\midrule
   $4$ &  $\begin{aligned}1 &+ 6 t^2 + 20 t^4 + 66 t^6 + 414 t^8 + 1332 t^{10} + 4046 t^{12} + 
 9743 t^{14} + 21420 t^{16} \\&+ 41790 t^{18} + 75726 t^{20} + 123425 t^{22} + 188793 t^{24} + 266485 t^{26} + 355729 t^{28} \\&+ 446666 t^{30} + 533544 t^{32} + 605586 t^{34} + 659290 t^{36} + 686624 t^{38} + \cdots + t^{78}\end{aligned}$\\\bottomrule
    \end{tabular}
    \caption{Numerator of unrefined Hilbert series of $i\sorm(N)$, $P_{12N+30}(t)$, for $N=0,1,2,3,4$.}
    \label{tab:P12N+30}
\end{table}

It is computationally unfeasible to extract the refined Hilbert series owing to the treatment of roots of unity in computer algebra systems
such as \texttt{Mathematica} or $\texttt{sympy}$ and hence intractable to compute the HWG. Given the complexity of the HWG of the $h\sorm(N)$ singularities, it is reasonable to expect that the HWG of the $i\sorm(N)$ singularities are at least as complicated.
\subsection{$g\sorm(N)/\mathbb Z_2$}
Just as with the special case of $N=2$ where $g\sorm(2)\simeq A_3$, there is an analogous result $g\sorm(2)/\mathbb Z_2\simeq A_7$ which is verified with the computation of the Higgs branch Hilbert series.

The more general cases of $g\sorm(N)/\mathbb Z_2$ for $N\geq 3$ share similar HWG which are related to that of $g\sorm(N)$ by a discrete quotient. The assignment of $\mathbb Z_2$ irreps in the HWG of $g\sorm(N)$ is summarised in Table \ref{tab:Z2gso(n)} together with the PL of the HWG of $g\sorm(N)/\mathbb Z_2$.
\begin{table}[h!]
    \centering
    
    \begin{tabular}{ccc}
    \toprule
     $N$& $\mathbb Z_2$ assignment on $\pl\left[\hwg\left[g\sorm(N)\right]\right]$ &  $\pl\left[\hwg\left[g\sorm(N)/\mathbb Z_2\right]\right]$\\\midrule
     $3$& $\mu^2 t^2 + (1+\epsilon\mu^6)t^4 + \epsilon\mu^6 t^6 - \mu^{12} t^{12}$& $\mu^2t^2+t^4+\mu^{12}t^8+\mu^{12}t^{10}-\mu^{24}t^{20}$ \\$4$& $(\mu_1^2 + \mu_2^2) t^2 + (1 + \epsilon\mu_1^3 \mu_2^3) t^4 + \epsilon\mu_1^3 \mu_2^3 t^6 - 
 \mu_1^6 \mu_2^6 t^{12}$ &$(\mu_1^2 + \mu_2^2) t^2 + t^4 + \mu_1^6 \mu_2^6 t^8 + \mu_1^6 \mu_2^6 t^{10} - \mu_1^{12} \mu_2^{12} t^{20}$\\$5$&$\mu_2^2t^2+(1+\mu_1^2+\epsilon\mu_1^3)t^4+\epsilon\mu_1^3t^6-\mu_1^6t^{12}$&$\mu_2^2 t^2 + (1+ \mu_1^2) t^4 + \mu_1^6 t^8 + \mu_1^6 t^{10} - \mu_1^{12} t^{20}$\\$6$&$\mu_2\mu_3t^2+(1+\mu_1^2+\epsilon\mu_1^3)t^4+\epsilon\mu_1^3t^6-\mu_1^6t^{12}$&$\mu_2 \mu_3 t^2 + (1 + \mu_1^2) t^4 + \mu_1^6 t^8 + \mu_1^6 t^{10} - \mu_1^{12} t^{20}$\\$\geq 7$&$\mu_2t^2+(1+\mu_1^2+\epsilon\mu_1^3)t^4+\epsilon\mu_1^3t^6-\mu_1^6t^{12}$&$\mu_2 t^2 + (1 + \mu_1^2) t^4 + \mu_1^6 t^8 + \mu_1^6 t^{10} - \mu_1^{12} t^{20}$\\\bottomrule
\end{tabular}

    \caption{$\mathbb Z_2$ representation assignments to terms in $\pl\left[\hwg\left[g\sorm(N)\right]\right]$ to produce the HWG of $g\sorm(N)/\mathbb Z_2$. The trivial representation is suppressed whilst the signed representation is denoted by $\epsilon$.}
    \label{tab:Z2gso(n)}
\end{table}
\subsection{$h\sorm(N)/\mathbb Z_2$}
A similar analysis can be done to compute Hilbert series and HWG for $h\sorm(N)/\mathbb Z_2$. First is the special case where $N=1$, it was found that $h\sorm(1)=E_6$ and so one should expect that $h\sorm(1)/\mathbb Z_2=E_6/\mathbb Z_2=E_7$, the Klein $E_7$ singularity. Indeed this is the case from computation of the Higgs branch Hilbert series of \Quiver{quiv:SymkZ2Family} with $k=5, N=1$. This suggests the following quotient constructions of the Klein $E_7$ singularity \begin{equation}
    \mathbb C^2/\Gamma_{E_7}=\mathbb C^{96}/\!/\!/\left[\urm(1)\times\urm(2)^3\times\urm(3)^2\times\urm(4)\right]=\mathbb C^8/\!/\!/\left[\sprm(1)\times\orm(1)\right]
\end{equation} The first being the discrete quotient of $\mathbb C^2$ binary octahedral group, the second is the Higgs branch of the affine $\widehat E_7$ quiver, and the third being the Higgs branch of \Quiver{quiv:SymkZ2Family} with $k=5, N=1$.

The more generic case with $N\geq3$ have similar HWG, which is again a product of a polynomial with the PE of a polynomial \begin{equation}
    \hwg\left[h\sorm(N)/\mathbb Z_2\right]=f'_N(t,\mu_1,\mu_2)\times\pe\left[g'_N(t,\mu_1,\mu_2,\mu_3)\right]
\end{equation}where the polynomials $f'_N(t,\mu_1,\mu_2)$ and $g'_N(t,\mu_1,\mu_2,\mu_3)$ are related to their un-primed counterparts in the HWG of $h\sorm(N)$ by the action of $\mathbb Z_2$. 

In the following, the explicit form of these polynomials and the action of the $\mathbb Z_2$ is presented. As before, the trivial representation of $\mathbb Z_2$ is suppressed and the non-trivial representation is denoted by $\epsilon$.

For the case of $N\geq 5$, the function $f'_{N}(t,\mu_1)$ is given as \begin{align}
    f'_{N\geq5}(t,\mu_1)&=1 + (\mu_1^2 + \mu_1^4) t^8 + (\mu_1^2 + \mu_1^4 + \mu_1^6) t^{10} + (\mu_1^2 + \mu_1^4 + 
    \mu_1^6 + \mu_1^8) t^{12} \nonumber\\&+ (\mu_1^2 + \mu_1^4 + \mu_1^6 + \mu_1^8 + 
    \mu_1^{10}) t^{14} + (1 + \mu_1^2 + \mu_1^4 + \mu_1^6 + \mu_1^8 + \mu_1^{10} + \mu_1^{12} + 
    \mu_1^{14}) t^{18} \nonumber\\&- (\mu_1^2 + \mu_1^4 + \mu_1^6 + \mu_1^8 + \mu_1^{10} + \mu_1^{12} + 
    \mu_1^{14} + \mu_1^{16}) t^{22} - (\mu_1^6 + \mu_1^8 + \mu_1^{10} + \mu_1^{12} + 
    \mu_1^{14}) t^{26} \nonumber\\&- (\mu_1^8 + \mu_1^{10} + \mu_1^{12} + \mu_1^{14}) t^{28} - (\mu_1^{10} + 
    \mu_1^{12} + \mu_1^{14}) t^{30} - (\mu_1^{12} + \mu_1^{14}) t^{32} - \mu_1^{16} t^{40}
\end{align}
The assignment of the $\mathbb Z_2$ representations for the un-primed polynomials in the HWG of $h\sorm(N)$ for $N\geq 5$ are 
\begin{align}
    f_{N\geq 5}(t,\mu)&=1 + \epsilon (\mu_1 + \mu_1^3) t^6 + \left(\mu_1^2 + \mu_1^4 + 
    \epsilon (\mu_1 + \mu_1^3 + \mu_1^{5})\right) t^8 + (\mu_1^2 + \mu_1^4 + \mu_1^{6}) t^{10} \nonumber\\&+ \left(\mu_1^2 + \mu_1^4 +
     \epsilon (\mu_1 + \mu_1^3 + \mu_1^{5} + \mu_1^{7} + \mu_1^{9})\right) t^{12} + (\mu_1^2 + \mu_1^4 + 
    \epsilon (\mu_1 + \mu_1^3 + \mu_1^{5})) t^{14} \nonumber\\&- 
 \epsilon (\mu_1^{7} + \mu_1^{9} + \mu_1^{11}) t^{16} + (1 + \mu_1^2 + \mu_1^4) t^{18} - \left(\mu_1^{6} + \mu_1^{8} +
     \mu_1^{10} + \epsilon (\mu_1^{7} + \mu_1^{9})\right) t^{20} \nonumber\\&- \left(\mu_1^2 + \mu_1^4 + \mu_1^{6} + \mu_1^{8} + 
    \mu_1^{10} + \epsilon (\mu_1^{7} + \mu_1^{9})\right) t^{22} - 
 \epsilon (\mu_1^{5} + \mu_1^{7} + \mu_1^{9}) t^{24} \nonumber\\&- \left(\mu_1^{6} + \mu_1^{8} + \mu_1^{10} + 
    \epsilon (\mu_1^{7} + \mu_1^{9})\right) t^{26} - (\mu_1^{8} + \mu_1^{10}) t^{28} - \epsilon \mu_1^{11} t^{34}
\end{align}The two formulae above also apply for the case of $N=3, 4$ with the replacement of $\mu_1\leftrightarrow \mu^2, \mu_1\mu_2$, respectively in each formula.

Similarly for $N\geq7$ the formula for $g'_{N}(t,\mu_1,\mu_2)$ is given as 
\begin{equation}
     g'_{N\geq7}(t,\mu_1,\mu_2)= \mu_2 t^2+ (1+\mu_1^2)t^4+ (1+\mu_1^{6})t^8 +(1+\mu_1^{10})t^{12}
\end{equation}with the assignment of $\mathbb Z_2$ representations for the un-primed polynomial $g_{N\geq7}(t,\mu_1,\mu_2)$ as \begin{equation}
    g_{N\geq 7}(t,\mu_1,\mu_2)=\mu_2 t^2 +(1+\mu_1^2) t^4+ \epsilon\mu_1^5 t^6+ (1+\mu_1^6)t^8 +t^{12}
\end{equation}the above two formulae are also valid for $N=5,6$ with the replacement of $\mu_2\leftrightarrow\mu_2^2, \mu_2\mu_3$, respectively.

However for the case of $N=3,4$ we have \begin{align}
    g'_3(t,\mu) &= \mu^2 t^2+ t^4+ (1+\mu^{12})t^8 +(1+\mu^{20})t^{12}\\
    g'_4(t,\mu_1,\mu_2)&= (\mu_1^2+\mu_2^2) t^2+ t^4+ (1+\mu_1^{6}\mu_2^6)t^8 +(1+\mu_1^{10}\mu_2^{10})t^{12}
\end{align}with the assignment of the $\mathbb Z_2$ representations as \begin{align}
    g_3(t,\mu)&=\mu^2 t^2+ t^4+\epsilon\mu^{{10}} t^6+  (1+\mu^{12})t^8 +t^{12}\\
    g_4(t,\mu_1,\mu_2)&= (\mu_1^2+\mu_2^2) t^2 +t^4+ \epsilon\mu_1^5\mu_2^5 t^6+ (1+\mu_1^6\mu_2^6)t^8 +t^{12}
\end{align}

\subsection{$i\sorm(N)/\mathbb Z_2$}
Once again it is challenging to extract refined Hilbert series or compute HWG due to the handling of seventh roots of unity in common programmes. It is possible to compute unrefined Hilbert series and extract some general features.

The case with $N=0$ gave $i\sorm(0)=E_8$, upon performing a $\mathbb Z_2$ quotient which is reflected in the Higgs branch of \Quiver{quiv:SymkZ2Family}, one finds Klein $E_8$ again. This suggests that the $\mathbb Z_2$ action as specified by the quiver is trivial on the Klein $E_8$ singularity.

In general for $N\geq 0$, the unrefined Hilbert series takes the following form \begin{equation}
    \hs\left[i\sorm(N)/\mathbb Z_2\right]=\frac{P_{26N+30}(t)}{(1-t^{12})^{N+1}(1-t^{16})^{N}(1-t^{20})}
\end{equation}where $P_{26N+30}(t)$ is a palindromial of degree $26N+30$. Some examples of $P_{26N+30}(t)$ are given in Table \ref{tab:P26N+30}.
\begin{table}[h!]
    \centering
    \begin{tabular}{cc}
    \toprule
        $N$ & $P_{26N+30}(t)$  \\
        \midrule
        $0$&$1+t^{30}$\\\midrule
        $1$ & $\begin{aligned}1 &+ 3 t^8 + 4 t^{12} + 4 t^{14} + 7 t^{16} + 9 t^{18} + 10 t^{20} + 13 t^{22} + 
 13 t^{24} + 18 t^{26} + 12 t^{28} \\&+ 18 t^{30} + 13 t^{32} + 13 t^{34} + 
 10 t^{36} + 9 t^{38} + 7 t^{40} + 4 t^{42} + 4 t^{44} + 3 t^{48} + t^{56}\end{aligned}$\\\midrule
         $2$& $\begin{aligned}1 &+ t^2 + t^4 + t^6 + 16 t^8 + 16 t^{10} + 49 t^{12} + 81 t^{14} + 
 151 t^{16} + 234 t^{18} \\&+ 357 t^{20} + 504 t^{22} + 698 t^{24} + 946 t^{26} + 
 1140 t^{28} + 1417 t^{30} \\&+ 1663 t^{32} + 1908 t^{34} + 2050 t^{36} + 
 2212 t^{38} + 2282 t^{40} + \cdots + t^{82}\end{aligned}$\\\midrule
         $3$ & $\begin{aligned}1 &+ 3 t^2 + 6 t^4 + 10 t^6 + 64 t^8 + 134 t^{10} + 377 t^{12} + 
 822 t^{14} + 1760 t^{16} + 3337 t^{18} \\&+ 6058 t^{20} + 10155 t^{22} + 16532 t^{24} + 25751 t^{26} + 37966 t^{28} + 54163 t^{30} + 74745 t^{32} \\&+ 99899 t^{34} + 128506 t^{36} + 161123 t^{38} + 196034 t^{40} + 232130 t^{42} + 267091 t^{44} \\&+ 299211 t^{46} + 327132 t^{48} + 348498 t^{50} + 361872 t^{52} + 365816 t^{54} + \cdots + t^{108}\end{aligned}$\\\midrule
   $4$ &  $\begin{aligned}1 &+ 6 t^2 + 20 t^4 + 50 t^6 + 234 t^8 + 696 t^{10} + 2168 t^{12} + 
 5751 t^{14} + 14286 t^{16} \\&+ 32042 t^{18} + 67390 t^{20} + 131581 t^{22} + 
 245250 t^{24} + 436021 t^{26} + 739461 t^{28} \\&+ 1203302 t^{30} + 
 1888815 t^{32} + 2865146 t^{34} + 4201730 t^{36} + 5977781 t^{38} \\&+ 
 8260216 t^{40} + 11105214 t^{42} + 14533946 t^{44} + 18540596 t^{46} + 
 23078015 t^{48} \\&+ 28053636 t^{50} + 33315872 t^{52} + 38669898 t^{54} + 
 43911871 t^{56} + 48792361 t^{58} \\&+ 53061669 t^{60} + 56486428 t^{62} + 
 58893221 t^{64} + 60135694 t^{66} + \cdots + t^{134}\end{aligned}$\\\bottomrule
    \end{tabular}
    \caption{Numerator of unrefined Hilbert series of $i\sorm(N)/\mathbb Z_2$, $P_{26N+30}(t)$, for $N=0,1,2,3,4$.}
    \label{tab:P26N+30}
\end{table}
\section{Coulomb Branch Hilbert Series}
\label{sec:CoulHS}
The magnetic quivers for the $g\sorm(N)$ can only identify one moduli space, the Coulomb branch, and not the Higgs branch owing to the non-simply laced edges. The electric quiver does not have this limitation and so the Higgs branch and Coulomb branch Hilbert series may be computed. The Higgs branch Hilbert series of \Quiver{quiv:SymkFamily} were computed previously in Section \ref{sec:HiggsBranchHS} corresponding to the $g\sorm(N),\;h\sorm(N),$ and $i\sorm(N)$ singularities. The Coulomb branches of \Quiver{quiv:SymkFamily} are symplectic duals of these singularities, this should be a one-dimensional space since the rank of the $\sprm(1)$ gauge group is one and therefore the Coulomb branch should be related in some close way to the Klein singularities.

It is important to recall that, for some parameters, namely when the quantity $\delta(k,N)$ defined in \eqref{e:delta-def} below is odd (equivalently, $N+\frac{k+1}{2}$ is odd),
the theory has a parity anomaly and so a formal definition of the Coulomb branch is not known. Nevertheless, one can compute a Hilbert series using the methods described below and study it as an object on its own. Only for the cases of even parameter
$\delta(k,N)$ below 
can the Hilbert series be associated to the Coulomb branch of the theory. Nevertheless we refer to such a Hilbert series as a ``Coulomb branch Hilbert series'' for both cases.

The Hilbert series, computed with the monopole formula \cite{Cremonesi:2013lqa}, sheds light on the identities of the Coulomb branch.
\begin{equation}
    \hs_{\mathcal C}\left[{\text{\Quiver{quiv:SymkFamily}}}\right]=\sum_{m=0}^\infty t^{2\Delta(m)}P_{\sprm(1)}(t,m)
\end{equation}where the conformal dimension $\Delta(m)$ is written in terms of a function of $k$ and $N$ called $\delta(k,N)$ whose expression is in the square brackets of the first equality \begin{equation}\label{e:delta-def}
    \Delta(m)=\frac{1}{2}|m|\left[N+\left(\frac{k+1}{2}\right)^2-4\right]=\frac{1}{2}\cdot|m|\cdot \delta(k,N)
\end{equation} and the dressing factor $P_{\sprm(1)}(t,m)$ is given as \begin{equation}
    P_{\sprm(1)}(t,m) =\begin{cases}\frac{1}{1-t^4},& m=0\\\frac{1}{1-t^2},&m\neq0\end{cases}
\end{equation}

The Coulomb branch Hilbert series may be evaluated exactly for all $k$ and $N$ as \begin{align}
    \hs_{\mathcal C}\left[{\text{\Quiver{quiv:SymkFamily}}}\right]&=\frac{1-t^{2\delta(k,N)+4}}{(1-t^4)(1-t^{\delta(k,N)})(1-t^{\delta(k,N)+2})}\\
    &=\pe\left[t^4+t^{\delta(k,N)}+t^{\delta(k,N)+2}-t^{2\delta(k,N)+4}\right]
\end{align}which suggests that the moduli space is a complete intersection.

If $\delta(k,N)$ is even, the moduli space can be identified as the Klein $D_{\delta(k,N)/2+2}$ singularity. If $\delta(k,N)$ is instead odd, then the Hilbert series is not that of a Klein singularity.
Note that the only quaternionic dimension one symplectic singularity is a Klein singularity, which means such Hilbert series does not reflect the algebra of a symplectic singularity.
Physically, this is an indication of the parity anomaly. However, on the level of the Hilbert series, there is still a $\mathbb{Z}_2$ action whose invariants produce a Hilbert series of a Klein singularity. Consider the general effect of assigning $\mathbb Z_2$ representations ($\mathbf{1}$ and $\epsilon$) to each of the generators in the following way and then summing over both conjugacy classes with the Molien average or Burnside lemma \cite{burnside_2012} \begin{equation}
    \pe\left[\mathbf1\cdot t^4+\epsilon\cdot t^{\delta(k,N)}+\epsilon\cdot t^{\delta(k,N)+2}-\mathbf{1}\cdot t^{2\delta(k,N)+4}\right]\rightarrow \pe\left[t^4+t^{2\delta(k,N)}+t^{2\delta(k,N)+2}- t^{4\delta(k,N)+4}\right]
\end{equation}which results in the Klein $D_{\delta(k,N)+2}$ singularity. Although this is true for all $\delta(k,N)$ this result is particularly useful when $\delta(k,N)$ is odd.
In summary \begin{equation}
    \hs_{\mathcal C}\left[{\text{\Quiver{quiv:SymkFamily}}}\right]=\begin{cases}
        \hs\left[D_{\delta(k,N)/2+2}\right],&\delta(k,N)\;\text{even}\\\mathbb Z_2\;\text{cover of}\;\hs\left[D_{\delta(k,N)+2}\right],&\delta(k,N)\;\text{odd}
    \end{cases}
\end{equation}
But we clarify that this $\mathbb Z_2$ cover of $D_{\delta(k,N)+2}$ for the $\delta(k,N)$ odd case is a statement at the level of the Hilbert series only. An honest homogeneous branched $\mathbb Z_2$ cover of a Klein singularity, if normal, must always produce another Klein singularity, because it would restrict to a covering over the smooth locus, which are classified by subgroups of the fundamental group, the corresponding finite subgroup of $\sprm(1)$ (or simply because all conical symplectic singularities of complex dimension two are Klein singularities). We note that there do exist non-normal cones with the Hilbert series for the odd $\delta(k,N)$ cases, though they are  not homogeneous branched  $\mathbb Z_2$ covers of type $D$ Klein singularities.

\subsection{Interesting Cases}
The discussion above about the Coulomb branch Hilbert series of \Quiver{quiv:SymkFamily} is quite general and applies for any $k$ and $N$. That being said, there are some interesting cases which deserve some further attention. These are the cases where the Higgs branch of \Quiver{quiv:SymkFamily} is a Klein singularity.

\paragraph{$\mathbf{k=3, N=2}$}
The Higgs branch of \Quiver{quiv:SymkFamily} in this case is Klein $A_3\simeq D_3$. As for the Coulomb branch, one computes $\delta(k,N) = 2$ which is even and so the Coulomb branch is also the Klein $A_3\simeq D_3$ singularity. Since the Higgs branch and the Coulomb are the same, one concludes that for $k=3,N=2$, \Quiver{quiv:SymkFamily} is self-mirror under $3d$ mirror symmetry \cite{Intriligator:1996ex}.

It should also be noted that this gauge theory can also be realised in six-dimensions from F-theory \cite{Klevers:2016jsz}.
\paragraph{$\mathbf{k=5, N=1}$}The Higgs branch of \Quiver{quiv:SymkFamily} is the Klein $E_6$ singularity. The value of $\delta(5,1)=6$ is even which means that the Coulomb branch is the Klein $D_5$ singularity. This is particularly interesting since there is a $\mathbb Z_2$ relationship between these two singularities as \begin{equation}
    D_5/\mathbb Z_2=E_6
\end{equation}

\paragraph{$\mathbf{k=7, N=0}$} The Higgs branch of \Quiver{quiv:SymkFamily} is Klein $E_8$. The value of $\delta(7,0)=12$ is even and so the Coulomb branch is the Klein $D_8$ singularity. This is not directly connected to the Klein $E_8$ singularity, although it is related to the Klein $E_7$ singularity via \begin{equation}
    D_8/\mathbb Z_2=E_7
\end{equation}
\section{Conclusion}
\label{sec:conc}
In this paper a two-parameter family of $\sprm(1)$ supersymmetric $3d\;\mathcal N=4$ gauge theories is studied. The parameter $N$ determines the number of fundamental half-hypermultiplets and the parameter $k$ refers to the $\mathrm{Sym}^k$ half-hypermultiplet where $k=1,3,5,7$.

The Higgs branches of these theories are families of isolated conical symplectic singularities, the case of $k=1$ being the minimal nilpotent orbit closure of $\sorm(N+1)$, with the case of $k=3,5,7$ being slightly more exotic spaces called the $g\sorm(N),\;h\sorm(N),$ and $i\sorm(N)$ singularities. The latter two have not appeared in the literature before. This introduces brand new transverse slices in the study of symplectic singularities and it would be interesting to search for physical or mathematical contexts in which they arise. An argument that these are the only isolated symplectic singularities is provided by the Higgs mechanism.

The one-dimensional members of these families are the $A_3\simeq g\sorm(2),\;E_6\simeq h\sorm(1),$ and $E_8\simeq i\sorm(0)$ Klein singularities. This gives novel hyper-Kähler quotient constructions of these moduli spaces as the Higgs branch of the $\sprm(1)$ gauge theories studied here. The Klein $E_7\simeq h\sorm(1)/\mathbb Z_2$ singularity was also computed as a hyper-Kähler quotient by studying the $\mathbb Z_2$ quotient of the Higgs branch with $k=5,\;N=1$. One way to view this $\sprm(1)$ hyper-Kähler quotient construction is as a ``lift'' of the usual definition of the Klein singularities. The typical definition, is through the discrete group action of an ADE subgroup $\Gamma_{ADE}\subset\sprm(1)$ acting on $\mathbb C^2$. This $\sprm(1)$ hyper-Kähler quotient instead quotients by the full $\sprm(1)$ group on $\mathbb C^8$ but encodes the action of the discrete $\Gamma_{ADE}$ subgroup through the matter representations in this gauge theory.

\begin{align}
    \mathbb C^2/\Gamma_{X}&=\mathbb C^8/\!/\!/\sprm(1),\qquad X\in\{A_3, E_6, E_8\}\\
    \mathbb C^2/\Gamma_{E_7}&=\mathbb C^8/\!/\!/[\sprm(1)\times\orm(1)]
\end{align}
This hyper-Kähler quotient construction should also be viewed as complementary to the construction of Kronheimer using the affine quivers.

One advantage of the $\sprm(1)$ gauge theory presented in this work is in the low rank of the gauge group, making it concise for computations. A hope would be to use the cases of these $\sprm(1)$ theories that have Higgs branches which are Klein singularities (especially of exceptional type) as input in studying string backgrounds. One possible application of this theory may be in extracting a Kähler potential and perhaps computing metrics of the singular space.

An open problem is in determining a magnetic quiver for the $h\sorm(N)$ and $i\sorm(N)$ singularities. The classification of isolated conical symplectic singularities in \cite{Bourget:2025wsp} extends to magnetic quivers which have only unitary gauge groups with bifundamental matter and possibly non-simply laced edges. Therefore the magnetic quivers for the $h\sorm(N)$ and $i\sorm(N)$ must lie outside of this class, possibly with other types of gauge groups (orthosymplectic or exceptional) and/or possibly with more exotic matter representations.

It is always a worthwhile task to understand brane constructions of gauge theories as there may be additional insight that one can gain. It is currently unclear how to realise the $\mathrm{Sym}^k$ representation in a brane system, for $k>2$, in either Type IIA or Type IIB, and possibly requires new objects or phenomena to explain. The computations in this work show that for $k=3,\;N=2$ one finds a gauge theory that is self-mirror under 3d mirror symmetry. This could be useful in determining a precise brane system, since a Type IIB brane construction for this theory would have to be symmetric with respect to $\mathrm{D}5$ and $\mathrm{NS}5$ branes (and related objects). From there, one can hope to generalise the brane system.

\acknowledgments
The work of AH, GK, and DL is partially supported by STFC Consolidated Grant ST/X000575/1. The work of GK is supported by STFC DTP research studentship grant ST/X508433/1. The work of TS is partially supported by grant UKRI2779.
We would like to thank Antoine Bourget and Quentin Lamouret for helpful discussions and collaboration at the very early stages of this project. TS would like to thank Gwyn Bellamy and Alastair Craw, whose collaboration on a related project had an important influence on this one.
\appendix
\section{The Non-normal Variety $m_1$}
\label{app_m1}
The non-normal variety $m_1$, of quaternionic dimension $1$, is the simplest example of a non-normal variety whose smooth locus is holomorphic symplectic \cite{vinberg1972class}. The simplest definition of $m_1$ starts with the ring of functions of $\mathbb C^2$ with the two generators removed. The Hilbert series and highest weight generating function describe this process\begin{align}
    \hs(m_1)&=\frac{1}{(1-a t)(1-t/a)}-(a+1/a)t \ , \\
    \hwg(m_1)&=\frac{1}{1-\mu t}-\mu t=\sum_{1 \neq n\in \mathbb{Z}_{\geq0}} (\mu t)^n=\pe[\mu^2t^2+\mu^3t^3-\mu^6t^6]\ ,
\end{align}
with $a$ and $\mu$ the fugacity and highest weight fugacity respectively of the $\sprm(1)$ symmetry acting on $m_1$.

The ring of functions of $m_1$ can now be specified, using coordinates $x$ and $y$ of $\mathbb C^2$ as:
\begin{equation}
\mathcal{R}_{m_1}=\mathbb C[x^2,xy,y^2,x^3,x^2y,xy^2,y^3],
    \label{eq_ringm}
\end{equation}
with $m_1=\text{Spec}\ \mathcal{R}_{m_1}$.

The variety $m_1$ appears as the fixed locus of $S_k\times S_{n-k}$ in $\mathbb C^{2n}/S_n$ when $k\neq n-k$. To see this, take the coordinates of $\mathbb C^{2n}$ as $(x_1,y_1,\cdots,x_n,y_n)$ and remove the free $\mathbb C^2$ by setting $\sum_ix_i=\sum_iy_i=0$. The invariant subspace fixed by the action of $S_k\times S_{n-k}$ is (up to conjugation),
\begin{equation}
V=(\underbrace{-(n-k) x, -(n-k) y, \cdots, -(n-k) x, -(n-k) y}_{k}, \underbrace{k x, k y, \cdots, k x, k y}_{n-k}).
\end{equation}
Restricting $S_n$ invariant functions on $\mathbb C_0^{2n}$ to $V$ gives the ring $\mathcal R_{m_1}$. Hence the corresponding stratum closure is $m_1$.

The above construction of $m_1$ may be generalised easily to any dimension. Namely taking the ring of functions of $\mathbb C^l,\;l\in \mathbb N^+$ and removing the $l$ generators. However, the holomorphic-symplectic property of the variety is preserved if and only if $l$ is even.

\section{Branching of \texorpdfstring{$[k]$}{[k]} to \(A\)-, \(D\)- and \(E\)-Type Finite Subgroups of \(\mathrm{SU}(2)\)}
\label{app_branching}

Let
\begin{equation}
[k]:=\operatorname{Sym}^k(\mathbb C^2)=\operatorname{Sym}^k([1])
\end{equation}
be the irreducible representation of \(\mathrm{SU}(2)\) with highest weight \(k\). Its dimension is
\begin{equation}
\dim [k]=k+1.
\end{equation}

Let
\begin{equation}
\Gamma\subset \mathrm{SU}(2)
\end{equation}
be a finite subgroup of \(A\)-, \(D\)- or \(E\)-type. The \(A\)-type groups are the cyclic groups
\begin{equation}
\Gamma=\mathbb Z_l,
\end{equation}
corresponding to the affine Dynkin diagram
\begin{equation}
\widehat A_{l-1}.
\end{equation}
The \(D\)-type groups are the binary dihedral groups
\begin{equation}
\Gamma=\operatorname{Dic}_l,
\end{equation}
corresponding to the affine Dynkin diagram
\begin{equation}
\widehat D_{l+2}.
\end{equation}
The \(E\)-type groups are
\begin{equation}
\Gamma_{E_6}=2T,\qquad \Gamma_{E_7}=2O,\qquad \Gamma_{E_8}=2I,
\end{equation}
corresponding to
\begin{equation}
\widehat E_6,\qquad \widehat E_7,\qquad \widehat E_8.
\end{equation}

Let
\begin{equation}
\tilde{\Gamma}=\{\sigma_0,\sigma_1,\dots,\sigma_s\}
\end{equation}
be the set of irreducible representations of \(\Gamma\), with
\begin{equation}
\sigma_0=1
\end{equation}
the trivial representation. The McKay correspondence identifies the irreducible representations in \(\tilde{\Gamma}\) with the nodes of the affine Dynkin diagram attached to \(\Gamma\). Using this McKay correspondence, the irreducible representation is relabelled as
\begin{equation}
\rho_{n,r},
\end{equation}
where \(n=\dim \rho_{n,r}\), and where \(r\) distinguishes different nodes with the same dimension. Under this convention, the trivial representation is identified by the null node and is written as
\begin{equation}
\rho_{1,0}=1.
\end{equation}

The \(D\)- and \(E\)-type subgroups always contain the central element
\begin{equation}
-I\in \mathrm{SU}(2).
\end{equation}
For \(A\)-type, the cyclic subgroup \(\mathbb Z_l\) contains \(-I\) precisely when \(l\) is even. Whenever \(-I\in\Gamma\), a parity constraint is imposed on the branching. On \([k]\), the element \(-I\) acts by
\begin{equation}
(-I)\big|_{[k]}=(-1)^k.
\end{equation}
On an irreducible representation \(\rho_{n,r}\) of \(\Gamma\), Schur's lemma implies that \(-I\) acts by a scalar:
\begin{equation}
(-I)\big|_{\rho_{n,r}}=\epsilon_{n,r}\operatorname{id}_{\rho_{n,r}},
\qquad
\epsilon_{n,r}\in\{+1,-1\}.
\end{equation}
Therefore, if \(-I\in\Gamma\), then \(\rho_{n,r}\) can appear in \([k]\big|_\Gamma\) only if
\begin{equation}
\epsilon_{n,r}=(-1)^k.
\end{equation}
Thus, in the presence of \(-I\), only irreps on which \(-I\) acts by \(+1\) can occur for even \(k\), and only irreps on which \(-I\) acts by \(-1\) can occur for odd \(k\).

\subsection*{McKay Matrix, Tensoring, and the Plethystic Exponential}

Let
\begin{equation}
\rho_{\mathrm{def}}:=\mathbb C^2\big|_\Gamma
\end{equation}
be the defining two-dimensional representation of \(\Gamma\). In the \(A\)-type case, \(\rho_{\mathrm{def}}\) is reducible; in the \(D\)- and \(E\)-type cases, it is irreducible.

Let
\begin{equation}
Q_0(\Gamma)
\end{equation}
denote the set of node labels \((n,r)\) after the McKay identification. For each node \((n,r)\in Q_0(\Gamma)\), the tensor product with the defining representation is decomposed as
\begin{equation}
\rho_{\mathrm{def}}\otimes \rho_{n,r}
=
\sum_{(m,s)\in Q_0(\Gamma)} A^\Gamma_{(m,s),(n,r)}\rho_{m,s}.
\end{equation}
The matrix
\begin{equation}
A^\Gamma=\left(A^\Gamma_{(m,s),(n,r)}\right)_{(m,s),(n,r)\in Q_0(\Gamma)}
\end{equation}
is the McKay matrix of \(\Gamma\). Equivalently, it is the adjacency matrix of the affine Dynkin diagram attached to \(\Gamma\). Thus the affine Dynkin diagram encodes the operation of tensoring by \(\rho_{\mathrm{def}}\).

The branching is written as
\begin{equation}
[k]\big|_\Gamma
=
\sum_{(n,r)\in Q_0(\Gamma)} a^\Gamma_{n,r}(k)\rho_{n,r}.
\label{eq_gammabranching1}
\end{equation}
For each node \((n,r)\in Q_0(\Gamma)\), the multiplicity generating function is defined by
\begin{equation}
P^\Gamma_{n,r}(t)
:=
\sum_{k=0}^{\infty}a^\Gamma_{n,r}(k)t^k.
\label{eq_gammabranching2}
\end{equation}
Equivalently, \(a^\Gamma_{n,r}(k)\) is the multiplicity of the quiver-node representation \(\rho_{n,r}\) in \([k]\big|_\Gamma\).

The plethystic exponential packages all symmetric powers of the defining representation:
\begin{equation}
\operatorname{PE}([1]t)
=
\sum_{k=0}^{\infty}[k]t^k.
\end{equation}

After restriction to \(\Gamma\), the defining representation becomes \(\rho_{\mathrm{def}}\), so
\begin{equation}
\operatorname{PE}(\rho_{\mathrm{def}}t)
=
\sum_{k=0}^{\infty}\left([k]\big|_\Gamma\right)t^k.
\end{equation}
Using \eqref{eq_gammabranching1} and \eqref{eq_gammabranching2}, this becomes
\begin{equation}
\operatorname{PE}(\rho_{\mathrm{def}}t)
=
\sum_{(n,r)\in Q_0(\Gamma)}P^\Gamma_{n,r}(t)\rho_{n,r}.
\end{equation}

For \(g\in\Gamma\), if the eigenvalues of \(\rho_{\mathrm{def}}(g)\) are \(\lambda\) and \(\lambda^{-1}\), then
\begin{equation}
\operatorname{PE}(\rho_{\mathrm{def}}t)(g)
=
\frac{1}{(1-\lambda t)(1-\lambda^{-1}t)}
=
\frac{1}{\det(1-t\rho_{\mathrm{def}}(g))}.
\end{equation}
Therefore, defining the character $\chi_{n,r}(g)=\mathrm{tr}\rho_{n,r}(g)$,
\begin{equation}
P^\Gamma_{n,r}(t)
=
\frac{1}{|\Gamma|}
\sum_{g\in\Gamma}
\frac{\overline{\chi_{n,r}(g)}}{\det(1-t\rho_{\mathrm{def}}(g))}.
\end{equation}

This plethystic exponential identity is equivalent to the McKay recursion. Indeed, the \(\mathrm{SU}(2)\) Clebsch--Gordan rule
\begin{equation}
[1]\otimes [k]\cong [k+1]+[k-1]
\end{equation}
implies
\begin{equation}
(1-t[1]+t^2)\operatorname{PE}([1]t)=1.
\end{equation}
After restriction to \(\Gamma\), this becomes
\begin{equation}
(1-t\rho_{\mathrm{def}}+t^2)\operatorname{PE}(\rho_{\mathrm{def}}t)=1.
\end{equation}
Using the expansion
\begin{equation}
\operatorname{PE}(\rho_{\mathrm{def}}t)
=
\sum_{(n,r)\in Q_0(\Gamma)}P^\Gamma_{n,r}(t)\rho_{n,r},
\end{equation}
and using the McKay matrix to encode tensoring by \(\rho_{\mathrm{def}}\), the finite linear system
\begin{equation}
\left(I-tA^\Gamma+t^2I\right)P^\Gamma(t)=e_{1,0}
\end{equation}
is obtained. Here \(P^\Gamma(t)\) is the column vector whose entries are \(P^\Gamma_{n,r}(t)\), ordered according to the chosen quiver-node labeling, and
\begin{equation}
    e_{1,0}=(1,0\dots,0)^T
\end{equation}
is the standard basis vector corresponding to the null node \(\rho_{1,0}=1\). Hence
\begin{equation}
P^\Gamma(t)
=
\left(I-tA^\Gamma+t^2I\right)^{-1}e_{1,0}.
\end{equation}

Thus all branching coefficients are obtained at once from the plethystic exponential, while the same coefficients are obtained recursively from the McKay recursion. Setting
\begin{equation}
b^\Gamma(k)
=
\left(a^\Gamma_{n,r}(k)\right)_{(n,r)\in Q_0(\Gamma)},
\end{equation}
the coefficient recursion is
\begin{align}
b^\Gamma(0)&=e_{1,0},\\
b^\Gamma(1)&=v_{\mathrm{def}},\\
b^\Gamma(k+1)&=A^\Gamma b^\Gamma(k)-b^\Gamma(k-1).
\end{align}
Here \(v_{\mathrm{def}}\) is the multiplicity vector of the defining representation:
\begin{equation}
\rho_{\mathrm{def}}
=
\sum_{(n,r)\in Q_0(\Gamma)}(v_{\mathrm{def}})_{n,r}\rho_{n,r}.
\end{equation}
For \(D\)- and \(E\)-type subgroups, \(\rho_{\mathrm{def}}\) is irreducible, and
\begin{equation}
    v_{\mathrm{def}}=(\dots,0,1,0,\dots)
\end{equation}
is a standard basis vector with the only non-trivial entry corresponding to the defining irrep.
For \(A\)-type subgroups, \(\rho_{\mathrm{def}}\) splits as a sum of two one-dimensional irrep.

\subsection*{McKay Quivers and Node Labeling}

The following diagrams fix the quiver-node labeling used in the branching formulas.

\subsubsection*{\(A\)-Type Quiver}

For \(\Gamma=\mathbb Z_l\), the McKay quiver is \(\widehat A_{l-1}\). Its nodes are
\begin{equation}
Q_0(\mathbb Z_l)=\{(1,r)\mid r\in\mathbb Z/l\mathbb Z\}.
\end{equation}
The corresponding irreducible representations are
\begin{equation}
\rho_{1,r},
\qquad
r\in\mathbb Z/l\mathbb Z.
\end{equation}

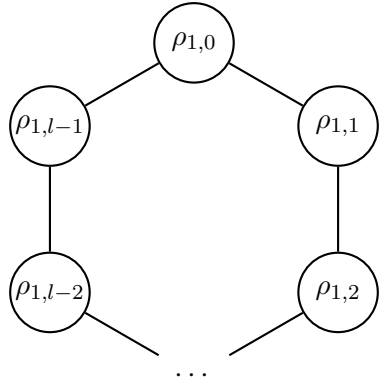
\begin{center}
\begin{tikzpicture}[
  scale=1.0,
  every node/.style={circle, draw, minimum size=1.05cm, inner sep=1pt},
  every path/.style={thick}
]
\node (r0) at (90:2.2cm) {$\rho_{1,0}$};
\node (r1) at (30:2.2cm) {$\rho_{1,1}$};
\node (r2) at (-30:2.2cm) {$\rho_{1,2}$};
\node[draw=none] (dots) at (-90:2.2cm) {$\cdots$};
\node (rl2) at (-150:2.2cm) {$\rho_{1,l-2}$};
\node (rl1) at (150:2.2cm) {$\rho_{1,l-1}$};

\draw (r0) -- (r1);
\draw (r1) -- (r2);
\draw (r2) -- (dots);
\draw (dots) -- (rl2);
\draw (rl2) -- (rl1);
\draw (rl1) -- (r0);
\end{tikzpicture}
\end{center}

The defining representation restricts as
\begin{equation}
\rho_{\mathrm{def}}
=
[1]\big|_{\mathbb Z_l}
=
\rho_{1,1}+\rho_{1,-1}.
\end{equation}
Thus tensoring by \(\rho_{\mathrm{def}}\) moves one step in either direction around the quiver:
\begin{equation}
\rho_{\mathrm{def}}\otimes\rho_{1,r}
=
\rho_{1,r+1}+\rho_{1,r-1}.
\end{equation}

\subsubsection*{\(D\)-Type Quiver}

For \(\Gamma=\operatorname{Dic}_l\), the McKay quiver is \(\widehat D_{l+2}\). Its nodes are
\begin{equation}
Q_0(\operatorname{Dic}_l)
=
\{(1,0),(1,1),(1,2),(1,3)\}
\cup
\{(2,j)\mid 1\le j\le l-1\}.
\end{equation}

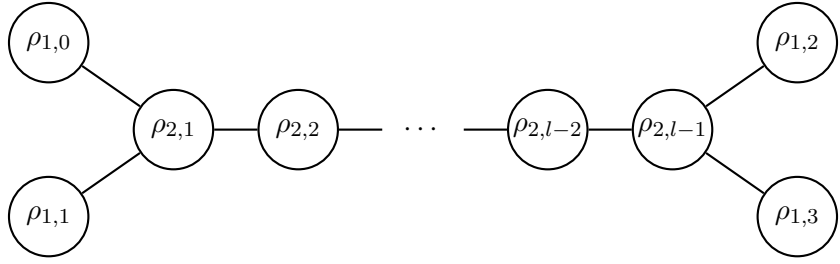
\begin{center}
\begin{tikzpicture}[
  x=1.65cm,
  y=1.15cm,
  every node/.style={circle, draw, minimum size=1.05cm, inner sep=1pt},
  every path/.style={thick}
]
\node (a0) at (0,1) {$\rho_{1,0}$};
\node (a1) at (0,-1) {$\rho_{1,1}$};
\node (b1) at (1,0) {$\rho_{2,1}$};
\node (b2) at (2,0) {$\rho_{2,2}$};
\node[draw=none] (dots) at (3,0) {$\cdots$};
\node (bl2) at (4,0) {$\rho_{2,l-2}$};
\node (bl1) at (5,0) {$\rho_{2,l-1}$};
\node (c2) at (6,1) {$\rho_{1,2}$};
\node (c3) at (6,-1) {$\rho_{1,3}$};

\draw (a0) -- (b1);
\draw (a1) -- (b1);
\draw (b1) -- (b2);
\draw (b2) -- (dots);
\draw (dots) -- (bl2);
\draw (bl2) -- (bl1);
\draw (bl1) -- (c2);
\draw (bl1) -- (c3);
\end{tikzpicture}
\end{center}

The defining representation is the node
\begin{equation}
\rho_{\mathrm{def}}
=
[1]\big|_{\operatorname{Dic}_l}
=
\rho_{2,1}.
\end{equation}
Thus tensoring by \(\rho_{2,1}\) is encoded by adjacency in the above \(\widehat D_{l+2}\) diagram.

\subsubsection*{\(E_6\)-Type Quiver}

For \(E_6=2T\), the McKay quiver is \(\widehat E_6\):

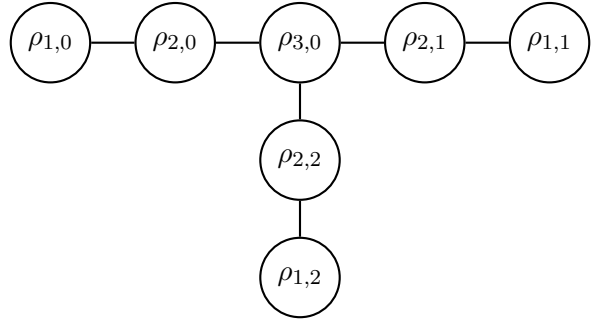
\begin{center}
\begin{tikzpicture}[
  x=1.65cm,
  y=1.15cm,
  every node/.style={circle, draw, minimum size=1.05cm, inner sep=1pt},
  every path/.style={thick}
]
\node (a) at (0,0) {$\rho_{1,0}$};
\node (b) at (1,0) {$\rho_{2,0}$};
\node (c) at (2,0) {$\rho_{3,0}$};
\node (d) at (3,0) {$\rho_{2,1}$};
\node (e) at (4,0) {$\rho_{1,1}$};
\node (f) at (2,-1.35) {$\rho_{2,2}$};
\node (g) at (2,-2.7) {$\rho_{1,2}$};

\draw (a) -- (b);
\draw (b) -- (c);
\draw (c) -- (d);
\draw (d) -- (e);
\draw (c) -- (f);
\draw (f) -- (g);
\end{tikzpicture}
\end{center}

Here the defining representation is
\begin{equation}
\rho_{\mathrm{def}}=\rho_{2,0}.
\end{equation}

\subsubsection*{\(E_7\)-Type Quiver}

For \(E_7=2O\), the McKay quiver is \(\widehat E_7\):

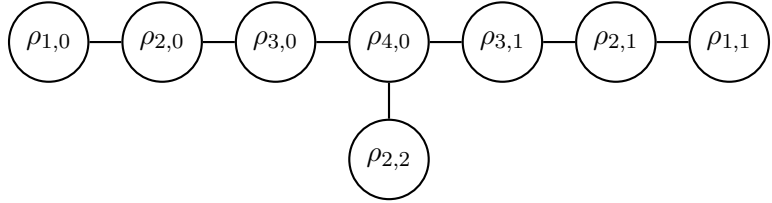
\begin{center}
\begin{tikzpicture}[
  x=1.5cm,
  y=1.15cm,
  every node/.style={circle, draw, minimum size=1.05cm, inner sep=1pt},
  every path/.style={thick}
]
\node (a) at (0,0) {$\rho_{1,0}$};
\node (b) at (1,0) {$\rho_{2,0}$};
\node (c) at (2,0) {$\rho_{3,0}$};
\node (d) at (3,0) {$\rho_{4,0}$};
\node (e) at (4,0) {$\rho_{3,1}$};
\node (f) at (5,0) {$\rho_{2,1}$};
\node (g) at (6,0) {$\rho_{1,1}$};
\node (h) at (3,-1.35) {$\rho_{2,2}$};

\draw (a) -- (b);
\draw (b) -- (c);
\draw (c) -- (d);
\draw (d) -- (e);
\draw (e) -- (f);
\draw (f) -- (g);
\draw (d) -- (h);
\end{tikzpicture}
\end{center}

Here the defining representation is
\begin{equation}
\rho_{\mathrm{def}}=\rho_{2,0}.
\end{equation}

\subsubsection*{\(E_8\)-Type Quiver}

For \(E_8=2I\), the McKay quiver is \(\widehat E_8\):

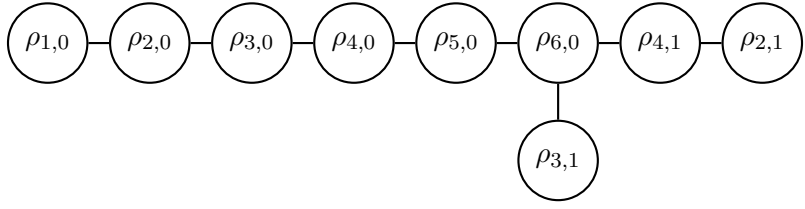
\begin{center}
\begin{tikzpicture}[
  x=1.35cm,
  y=1.15cm,
  every node/.style={circle, draw, minimum size=1.05cm, inner sep=1pt},
  every path/.style={thick}
]
\node (a) at (0,0) {$\rho_{1,0}$};
\node (b) at (1,0) {$\rho_{2,0}$};
\node (c) at (2,0) {$\rho_{3,0}$};
\node (d) at (3,0) {$\rho_{4,0}$};
\node (e) at (4,0) {$\rho_{5,0}$};
\node (f) at (5,0) {$\rho_{6,0}$};
\node (g) at (6,0) {$\rho_{4,1}$};
\node (h) at (7,0) {$\rho_{2,1}$};
\node (i) at (5,-1.35) {$\rho_{3,1}$};

\draw (a) -- (b);
\draw (b) -- (c);
\draw (c) -- (d);
\draw (d) -- (e);
\draw (e) -- (f);
\draw (f) -- (g);
\draw (g) -- (h);
\draw (f) -- (i);
\end{tikzpicture}
\end{center}

Here the defining representation is
\begin{equation}
\rho_{\mathrm{def}}=\rho_{2,0}.
\end{equation}

\subsection*{\(A\)-Type Branching: \(\mathbb Z_l\)}

Let
\begin{equation}
\mathbb Z_l=\langle a\mid a^l=1\rangle
\end{equation}
be the cyclic subgroup of \(\mathrm{SU}(2)\) generated by
\begin{equation}
a=
\begin{pmatrix}
\omega & 0\\
0 & \omega^{-1}
\end{pmatrix},
\qquad
\omega^l=1.
\end{equation}

The irreducible representations are the one-dimensional representations
\begin{equation}
\rho_{1,r},
\qquad
r\in\mathbb Z/l\mathbb Z,
\end{equation}
defined by
\begin{equation}
\rho_{1,r}(a)=\omega^r.
\end{equation}
The defining representation restricts as
\begin{equation}
\rho_{\mathrm{def}}
=
[1]\big|_{\mathbb Z_l}
=
\rho_{1,1}+\rho_{1,-1}.
\end{equation}
Therefore
\begin{equation}
\rho_{\mathrm{def}}\otimes\rho_{1,r}
=
\rho_{1,r+1}+\rho_{1,r-1},
\end{equation}
so the McKay matrix is
\begin{equation}
A^{A_{l-1}}_{(1,s),(1,r)}
=
\delta_{s,r+1}+\delta_{s,r-1},
\qquad
r,s\in\mathbb Z/l\mathbb Z.
\end{equation}

The branching is written as
\begin{equation}
[k]\big|_{\mathbb Z_l}
=
\sum_{r\in\mathbb Z/l\mathbb Z}
a^{A_{l-1}}_{1,r}(k)\rho_{1,r}.
\end{equation}
The generating functions are defined by
\begin{equation}
P^{A_{l-1}}_{1,r}(t)
:=
\sum_{k=0}^{\infty}a^{A_{l-1}}_{1,r}(k)t^k.
\end{equation}

Using the plethystic exponential identity, one obtains
\begin{equation}
\operatorname{PE}\left((\rho_{1,1}+\rho_{1,-1})t\right)
=
\sum_{r\in\mathbb Z/l\mathbb Z}
P^{A_{l-1}}_{1,r}(t)\rho_{1,r}.
\end{equation}
Equivalently,
\begin{equation}
P^{A_{l-1}}_{1,r}(t)
=
\sum_{\substack{p,q\ge 0\\ p-q\equiv r\!\!\!\pmod l}}
t^{p+q}.
\end{equation}
A closed rational form is
\begin{equation}
P^{A_{l-1}}_{1,r}(t)
=
\frac{t^r+t^{l-r}}{(1-t^2)(1-t^l)},
\qquad
1\le r\le l-1,
\end{equation}
and
\begin{equation}
P^{A_{l-1}}_{1,0}(t)
=
\frac{1+t^l}{(1-t^2)(1-t^l)}.
\end{equation}
Hence one obtains
\begin{equation}
\operatorname{PE}\left([1]\vert_{\mathbb Z_l} t\right)
=
\frac{1}{(1-t^2)(1-t^l)}\left(
(1+t^l)\rho_{1,0}
+
\sum_{r=1}^{l-1}(t^r+t^{l-r})\rho_{1,r}\right).
\end{equation}
Therefore
\begin{equation}
[k]\big|_{\mathbb Z_l}
=
\sum_{r\in\mathbb Z/l\mathbb Z}
a^{A_{l-1}}_{1,r}(k)\rho_{1,r}.
\end{equation}
Equivalently, the weights of \([k]\) are
\begin{equation}
k,\ k-2,\ k-4,\ \dots,\ -k.
\end{equation}
Thus
\begin{equation}
a^{A_{l-1}}_{1,r}(k)
=
\#\left\{
s\in\{0,1,\dots,k\}
\mid
k-2s\equiv r\pmod l
\right\}.
\end{equation}

\subsection*{\(D\)-Type Branching: \(\operatorname{Dic}_l\)}

Let
\begin{equation}
\operatorname{Dic}_l
=
\langle a,b\mid a^{2l}=1,\ b^2=a^l,\ bab^{-1}=a^{-1}\rangle.
\end{equation}

Its irreducible representations are labelled by the nodes of \(\widehat D_{l+2}\):
\begin{equation}
Q_0(\operatorname{Dic}_l)
=
\{(1,0),(1,1),(1,2),(1,3)\}
\cup
\{(2,j)\mid 1\le j\le l-1\}.
\end{equation}
The four one-dimensional representations are
\begin{equation}
\rho_{1,0},\qquad
\rho_{1,1},\qquad
\rho_{1,2},\qquad
\rho_{1,3},
\end{equation}
with
\begin{equation}
\rho_{1,0}=1.
\end{equation}
The remaining irreducible representations are the \(l-1\) two-dimensional representations
\begin{equation}
\rho_{2,j},
\qquad
1\le j\le l-1.
\end{equation}

The defining representation is
\begin{equation}
\rho_{\mathrm{def}}
=
[1]\big|_{\operatorname{Dic}_l}
=
\rho_{2,1}.
\end{equation}
With respect to the quiver labeling above, the tensor-product rules are
\begin{align}
\rho_{2,1}\otimes \rho_{1,0}&=\rho_{2,1},\\
\rho_{2,1}\otimes \rho_{1,1}&=\rho_{2,1},\\
\rho_{2,1}\otimes \rho_{1,2}&=\rho_{2,l-1},\\
\rho_{2,1}\otimes \rho_{1,3}&=\rho_{2,l-1},
\end{align}
and
\begin{align}
\rho_{2,1}\otimes \rho_{2,1}
&=
\rho_{1,0}+\rho_{1,1}+\rho_{2,2},\\
\rho_{2,1}\otimes \rho_{2,j}
&=
\rho_{2,j-1}+\rho_{2,j+1},
\qquad 2\le j\le l-2,\\
\rho_{2,1}\otimes \rho_{2,l-1}
&=
\rho_{2,l-2}+\rho_{1,2}+\rho_{1,3}.
\end{align}
Thus the McKay matrix is the adjacency matrix of \(\widehat D_{l+2}\) with the node labeling drawn above.

The branching is written as
\begin{align}
[k]\big|_{\operatorname{Dic}_l}
&=
a^{D_l}_{1,0}(k)\rho_{1,0}
+
a^{D_l}_{1,1}(k)\rho_{1,1}
+
a^{D_l}_{1,2}(k)\rho_{1,2}
+
a^{D_l}_{1,3}(k)\rho_{1,3} \notag\\
&\qquad
+
\sum_{j=1}^{l-1}a^{D_l}_{2,j}(k)\rho_{2,j}.
\end{align}where the $l$ is the dicyclic parameter rather than a Dynkin label.

The generating functions are defined by
\begin{equation}
P^{D_l}_{n,r}(t)
=
\sum_{k=0}^{\infty}a^{D_l}_{n,r}(k)t^k.
\end{equation}
Set
\begin{equation}
D_{D_l}(t):=(1-t^2)(1-t^{2l}).
\end{equation}

The two-dimensional nodes have generating functions
\begin{equation}
P^{D_l}_{2,j}(t)
=
\frac{t^j+t^{2l-j}}{D_{D_l}(t)},
\qquad
1\le j\le l-1.
\end{equation}
The one-dimensional nodes have generating functions
\begin{align}
P^{D_l}_{1,0}(t)
&=
\frac{1+t^{2l+2}}{(1-t^4)(1-t^{2l})},\\
P^{D_l}_{1,1}(t)
&=
\frac{t^2+t^{2l}}{(1-t^4)(1-t^{2l})},\\
P^{D_l}_{1,2}(t)
&=
\frac{t^l}{D_{D_l}(t)},\\
P^{D_l}_{1,3}(t)
&=
\frac{t^l}{D_{D_l}(t)}.
\end{align}
Hence one obtains the generating function of all irreps of $\surm(2)$ branching into irreps of $\mathrm{Dic}_l$
\begin{align}
\operatorname{PE}([1]\vert_{\mathrm{Dic}_l}t)
&=
\frac{1}{(1-t^4)(1-t^{2l})}\Bigg(
(1+t^{2l+2})\rho_{1,0}
+
(t^2+t^{2l})\rho_{1,1}
\notag\\
&\qquad
+
t^l(1+t^2)(\rho_{1,2}+\rho_{1,3})
+
(1+t^2)\sum_{j=1}^{l-1}(t^j+t^{2l-j})\rho_{2,j}\Bigg).
\end{align}
The branching coefficients \(a^{D_l}_{n,r}(k)\) are defined by these generating functions.

\subsection*{\(E\)-Type Branching: \(E_6,E_7,E_8\)}

For \(E\)-type, the representation labels are chosen so that the McKay quiver is written directly in terms of the nodes \(\rho_{n,r}\). In each case, \(\rho_{1,0}=1\) is the trivial representation and \(\rho_{2,0}\) is the defining representation.

\subsubsection*{\(E_6=2T\)}

For \(E_6=2T\), the node set is
\begin{equation}
Q_0(E_6)
=
\{(1,0),(2,0),(3,0),(2,1),(1,1),(2,2),(1,2)\}.
\end{equation}
The defining representation is
\begin{equation}
\rho_{\mathrm{def}}=\rho_{2,0}.
\end{equation}
Set
\begin{equation}
D_{E_6}(t):=(1-t^4)(1-t^6).
\end{equation}

The branching is written as
\begin{equation}
[k]\big|_{E_6}
=
\sum_{(n,r)\in Q_0(E_6)}a^{E_6}_{n,r}(k)\rho_{n,r}.
\end{equation}
The generating functions are defined by
\begin{equation}
P^{E_6}_{n,r}(t)
=
\sum_{k=0}^{\infty}a^{E_6}_{n,r}(k)t^k.
\end{equation}

The multiplicity generating functions are
\begin{align}
P^{E_6}_{1,0}(t)
&=
\frac{1-t^4+t^8}{D_{E_6}(t)},&
P^{E_6}_{2,0}(t)
&=
\frac{t+t^7}{D_{E_6}(t)},\\
P^{E_6}_{3,0}(t)
&=
\frac{t^2+t^4+t^6}{D_{E_6}(t)},&
P^{E_6}_{2,1}(t)
&=
\frac{t^3+t^5}{D_{E_6}(t)},\\
P^{E_6}_{2,2}(t)
&=
\frac{t^3+t^5}{D_{E_6}(t)},&
P^{E_6}_{1,1}(t)
&=
\frac{t^4}{D_{E_6}(t)},\\
P^{E_6}_{1,2}(t)
&=
\frac{t^4}{D_{E_6}(t)}.
\end{align}
Hence one obtains the generating function of all irreps of $SU(2)$ branching into irreps of $2T$
\begin{align}
\operatorname{PE}([1]\vert_{E_6}t)
&=
\frac{1}{(1-t^4)(1-t^6)}\Bigg(
(1-t^4+t^8)\rho_{1,0}
+
(t+t^7)\rho_{2,0}
+
(t^2+t^4+t^6)\rho_{3,0}
\notag\\
&\qquad
+
(t^3+t^5)\rho_{2,1}
+
(t^3+t^5)\rho_{2,2}
+
t^4\rho_{1,1}
+
t^4\rho_{1,2}\Bigg).
\end{align}

For even \(k\),
\begin{equation}
[k]\big|_{E_6}
=
a^{E_6}_{1,0}(k)\rho_{1,0}
+
a^{E_6}_{3,0}(k)\rho_{3,0}
+
a^{E_6}_{1,1}(k)\rho_{1,1}
+
a^{E_6}_{1,2}(k)\rho_{1,2}.
\end{equation}
For odd \(k\),
\begin{equation}
[k]\big|_{E_6}
=
a^{E_6}_{2,0}(k)\rho_{2,0}
+
a^{E_6}_{2,1}(k)\rho_{2,1}
+
a^{E_6}_{2,2}(k)\rho_{2,2}.
\end{equation}

\subsubsection*{\(E_7=2O\)}

For \(E_7=2O\), the node set is
\begin{equation}
Q_0(E_7)
=
\{(1,0),(2,0),(3,0),(4,0),(3,1),(2,1),(1,1),(2,2)\}.
\end{equation}
The defining representation is
\begin{equation}
\rho_{\mathrm{def}}=\rho_{2,0}.
\end{equation}
Set
\begin{equation}
D_{E_7}(t):=(1-t^6)(1-t^8).
\end{equation}

The branching is written as
\begin{equation}
[k]\big|_{E_7}
=
\sum_{(n,r)\in Q_0(E_7)}a^{E_7}_{n,r}(k)\rho_{n,r}.
\end{equation}
The generating functions are defined by
\begin{equation}
P^{E_7}_{n,r}(t)
=
\sum_{k=0}^{\infty}a^{E_7}_{n,r}(k)t^k.
\end{equation}

The multiplicity generating functions are
\begin{align}
P^{E_7}_{1,0}(t)
&=
\frac{1-t^6+t^{12}}{D_{E_7}(t)},&
P^{E_7}_{2,0}(t)
&=
\frac{t+t^{11}}{D_{E_7}(t)},\\
P^{E_7}_{3,0}(t)
&=
\frac{t^2+t^6+t^{10}}{D_{E_7}(t)},&
P^{E_7}_{4,0}(t)
&=
\frac{t^3+t^5+t^7+t^9}{D_{E_7}(t)},\\
P^{E_7}_{3,1}(t)
&=
\frac{t^4+t^6+t^8}{D_{E_7}(t)},&
P^{E_7}_{2,1}(t)
&=
\frac{t^5+t^7}{D_{E_7}(t)},\\
P^{E_7}_{1,1}(t)
&=
\frac{t^6}{D_{E_7}(t)},&
P^{E_7}_{2,2}(t)
&=
\frac{t^4+t^8}{D_{E_7}(t)}.
\end{align}
Hence one obtains the generating function of all irreps of $SU(2)$ branching into irreps of $2O$
\begin{align}
\operatorname{PE}([1]\vert_{E_7}t)
&=
\frac{1}{(1-t^6)(1-t^8)}\Bigg(
(1-t^6+t^{12})\rho_{1,0}
+
(t+t^{11})\rho_{2,0}
+
(t^2+t^6+t^{10})\rho_{3,0}
\notag\\
&\qquad
+
(t^3+t^5+t^7+t^9)\rho_{4,0}
+
(t^4+t^6+t^8)\rho_{3,1}
\notag\\
&\qquad
+
(t^5+t^7)\rho_{2,1}
+
t^6\rho_{1,1}
+
(t^4+t^8)\rho_{2,2}\Bigg).
\end{align}

For even \(k\),
\begin{align}
[k]\big|_{E_7}
&=
a^{E_7}_{1,0}(k)\rho_{1,0}
+
a^{E_7}_{3,0}(k)\rho_{3,0}
+
a^{E_7}_{3,1}(k)\rho_{3,1} \notag\\
&\qquad
+
a^{E_7}_{1,1}(k)\rho_{1,1}
+
a^{E_7}_{2,2}(k)\rho_{2,2}.
\end{align}
For odd \(k\),
\begin{equation}
[k]\big|_{E_7}
=
a^{E_7}_{2,0}(k)\rho_{2,0}
+
a^{E_7}_{4,0}(k)\rho_{4,0}
+
a^{E_7}_{2,1}(k)\rho_{2,1}.
\end{equation}

\subsubsection*{\(E_8=2I\)}

For \(E_8=2I\), the node set is
\begin{equation}
Q_0(E_8)
=
\{(1,0),(2,0),(3,0),(4,0),(5,0),(6,0),(4,1),(2,1),(3,1)\}.
\end{equation}
The defining representation is
\begin{equation}
\rho_{\mathrm{def}}=\rho_{2,0}.
\end{equation}
Set
\begin{equation}
D_{E_8}(t):=(1-t^{10})(1-t^{12}).
\end{equation}

The branching is written as
\begin{equation}
[k]\big|_{E_8}
=
\sum_{(n,r)\in Q_0(E_8)}a^{E_8}_{n,r}(k)\rho_{n,r}.
\end{equation}
The generating functions are defined by
\begin{equation}
P^{E_8}_{n,r}(t)
=
\sum_{k=0}^{\infty}a^{E_8}_{n,r}(k)t^k.
\end{equation}

The multiplicity generating functions are
\begin{align}
P^{E_8}_{1,0}(t)
&=
\frac{1-t^{10}+t^{20}}{D_{E_8}(t)},&
P^{E_8}_{2,0}(t)
&=
\frac{t+t^{19}}{D_{E_8}(t)},\\
P^{E_8}_{3,0}(t)
&=
\frac{t^2+t^{10}+t^{18}}{D_{E_8}(t)},&
P^{E_8}_{4,0}(t)
&=
\frac{t^3+t^9+t^{11}+t^{17}}{D_{E_8}(t)},\\
P^{E_8}_{5,0}(t)
&=
\frac{t^4+t^8+t^{10}+t^{12}+t^{16}}{D_{E_8}(t)},&
P^{E_8}_{6,0}(t)
&=
\frac{t^5+t^7+t^9+t^{11}+t^{13}+t^{15}}{D_{E_8}(t)},\\
P^{E_8}_{4,1}(t)
&=
\frac{t^6+t^8+t^{12}+t^{14}}{D_{E_8}(t)},&
P^{E_8}_{2,1}(t)
&=
\frac{t^7+t^{13}}{D_{E_8}(t)},\\
P^{E_8}_{3,1}(t)
&=
\frac{t^6+t^{10}+t^{14}}{D_{E_8}(t)}.
\end{align}
Hence one obtains the generating function of all irreps of $SU(2)$ branching into irreps of $2I$
\begin{align}
\operatorname{PE}([1]\vert_{E_8}t)
&=
\frac{1}{(1-t^{10})(1-t^{12})}\Bigg(
(1-t^{10}+t^{20})\rho_{1,0}
+
(t+t^{19})\rho_{2,0}
+
(t^2+t^{10}+t^{18})\rho_{3,0}
\notag\\
&\qquad
+
(t^3+t^9+t^{11}+t^{17})\rho_{4,0}
+
(t^4+t^8+t^{10}+t^{12}+t^{16})\rho_{5,0}
\notag\\
&\qquad
+
(t^5+t^7+t^9+t^{11}+t^{13}+t^{15})\rho_{6,0}
+
(t^6+t^8+t^{12}+t^{14})\rho_{4,1}
\notag\\
&\qquad
+
(t^7+t^{13})\rho_{2,1}
+
(t^6+t^{10}+t^{14})\rho_{3,1}\Bigg).
\end{align}

For even \(k\),
\begin{align}
[k]\big|_{E_8}
&=
a^{E_8}_{1,0}(k)\rho_{1,0}
+
a^{E_8}_{3,0}(k)\rho_{3,0}
+
a^{E_8}_{5,0}(k)\rho_{5,0} \notag\\
&\qquad
+
a^{E_8}_{4,1}(k)\rho_{4,1}
+
a^{E_8}_{3,1}(k)\rho_{3,1}.
\end{align}
For odd \(k\),
\begin{equation}
[k]\big|_{E_8}
=
a^{E_8}_{2,0}(k)\rho_{2,0}
+
a^{E_8}_{4,0}(k)\rho_{4,0}
+
a^{E_8}_{6,0}(k)\rho_{6,0}
+
a^{E_8}_{2,1}(k)\rho_{2,1}.
\end{equation}

\bibliographystyle{JHEP}
\bibliography{references}
\end{document}